\documentclass[graybox,vecphys]{svmult}

\usepackage[T1]{fontenc}
\usepackage{type1cm}        
\usepackage{makeidx}         
\usepackage{graphicx}        
\usepackage{multicol}        
\usepackage[bottom]{footmisc}

\usepackage{newtxtext}       %
\usepackage[varvw]{newtxmath}  
\usepackage{amsmath}
\usepackage{amsopn}
\usepackage{csquotes} 
\usepackage{color, xcolor}
\usepackage{hyperref}
\usepackage[capitalize]{cleveref}
\usepackage{siunitx}
\usepackage{tikz}
\usepackage{subcaption}
\usepackage{array} 

\usepackage{booktabs}

\newcommand{\pdifft}[1]{\frac{\partial  #1}{\partial t}}

\newcommand{\xx}{\vec{x}}
\newcommand{\nn}{\vec{n}}
\newcommand{\JJ}{\vec{J}}

\newcommand{\dx}{\, \mathrm{d}x}
\newcommand{\ds}{\, \mathrm{d}s}

\makeindex             

\begin{document}

\title*{Modeling and simulation of electrodiffusion in dense reconstructions of cerebral tissue}
\titlerunning{Electrodiffusion in dense tissue reconstructions}
\author{Halvor~Herlyng\orcidID{0000-0003-2301-9276}
    and\\ Marius~Causemann\orcidID{0000-0002-8579-2777}
    and\\ Gaute~T.~Einevoll\orcidID{0000-0002-5425-5012}
    and\\ Ada~J.~Ellingsrud\orcidID{0000-0002-8600-578X}
    and\\ Geir~Halnes\orcidID{0000-0002-4721-1599}
    and\\ Marie~E.~Rognes\orcidID{0000-0002-6872-3710}
}
\authorrunning{Herlyng~\emph{et al.}} 
\institute{
    Halvor~Herlyng, Marius~Causemann, Ada~J.~Ellingsrud \at Simula Research Laboratory, Kristian Augusts gate 23, 0164 Oslo, Norway, \email{hherlyng@simula.no, mariusca@simula.no, ada@simula.no} \and
    Marie~E.~Rognes \at Simula Research Laboratory, Kristian Augusts gate 23, 0164 Oslo, Norway and K. G. Jebsen Centre for Brain Fluid Research, Oslo, Norway \email{meg@simula.no} \and
    Gaute~T.~Einevoll \at Norwegian University of Life Sciences, \r{A}s, Norway and University of Oslo, Oslo, Norway \email{gaute.einevoll@nmbu.no} \and
    Geir~Halnes \at Norwegian Artificial Intelligence Research Consortium, Oslo, Norway, \email{geir.halnes@nora.ai}
}
\maketitle

\abstract{ Excitable tissue is fundamental to brain function, yet its
  study is complicated by extreme morphological complexity and the
  physiological processes governing its dynamics. Consequently,
  detailed computational modeling of this tissue represents a
  formidable task, requiring both efficient numerical methods and
  robust implementations. Meanwhile, efficient and robust methods for
  image segmentation and meshing are needed to provide realistic
  geometries for which numerical solutions are tractable. Here, we
  present a computational framework that models electrodiffusion in
  excitable cerebral tissue, together with realistic geometries
  generated from electron microscopy data. To demonstrate a possible
  application of the framework, we simulate electrodiffusive dynamics
  in cerebral tissue during neuronal activity. Our results and
  findings highlight the numerical and computational challenges
  associated with modeling and simulation of electrodiffusion and
  other multiphysics in dense reconstructions of cerebral tissue. }

\section{Introduction} 

Our brains are composed of complex cellular tissue consisting of nerve
and glial cells, separated by the tortuous extracellular space, and
with vasculature running throughout~\cite{kandel2000principles}. Brain
tissue is excitable: cells communicate through electromagnetic,
chemical and mechanical signals, with ion dynamics playing a central
role. Signals can be sent directly from cell to cell through synapses
via electric potential spikes known as action potentials (APs), while
electrochemical changes in the extracellular environment may also
excite or inhibit signals.

\begin{figure}
    \centering
    \includegraphics[width=\textwidth]{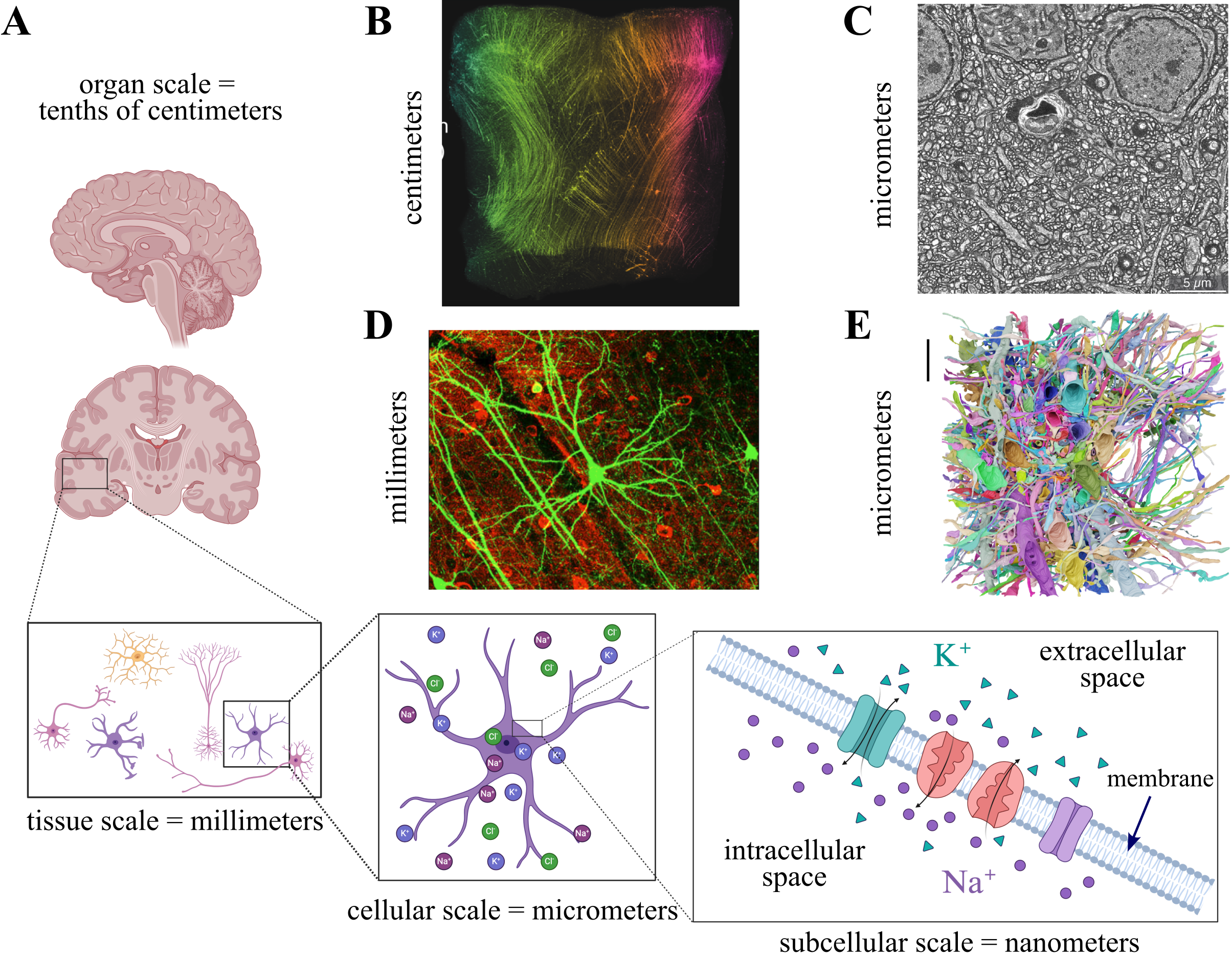}
    \caption{\textbf{A.} The physical scales of the brain, from organ
      scale to subcellular scale. At the subcellular scale, ions are
      transported across the cellular membrane through ion
      channels. The ion channels illustrated on the membrane are, from
      left to right: open generic ion channel,
      $\mathrm{Na}^+/\mathrm{K^+}$-ATPase pump importing potassium
      ions, $\mathrm{Na}^+/\mathrm{K^+}$-ATPase pump exporting sodium
      ions, and a closed generic ion channel. The illustrations in
      this panel were created using
      \href{https://BioRender.com/gc2jbqv}{Biorender}. \textbf{B.}
      Plane illumination microscopy of macaque primary motor cortex
      tissue at centimeter scale~\cite{glaser2025expansion}
      \textbf{C.} Confocal microscopy image of a neuron (green) in
      rodent visual cortex tissue. Image spans half a
      millimeter~\cite{lee2006dynamic}. \textbf{D.} Light-microscopy
      based reconstructed cross section of mouse hippocampus tissue at
      the micrometer scale~\cite{tavakoli2025light}. \textbf{E.} 3D
      rendering of mouse hippocampus tissue, scale bar 3
      $\mu$m~\cite{tavakoli2025light}. The images in panels
      \textbf{B--E} are reproduced under the terms of the
      \href{https://creativecommons.org/licenses/by/4.0/}{Creative
        Commons Attribution License}.}
    \label{fig:brain_scales}
\end{figure}

Understanding the physical mechanisms underlying brain function is key
to better understand neurodegenerative and order neurological
disorders. A 2024 study estimated that over one third of the global
population suffers or has suffered from a neurological condition, such
as stroke, epilepsy and dementia~\cite{steinmetz2024global}.
Computational modeling of neuronal biophysics offers a route to
improved understanding: models can be tailored to study specific
physical mechanisms, such as the response of a cell to an action
potential, or how changes in action potential firing frequencies
affect ion concentrations in the surrounding cerebral
tissue. Moreover, simulations are flexible because the choice of
parameters, initial and boundary conditions can often be altered
swiftly at low or no cost. However, an inherent modelling challenge is
the range of spatiotemporal scales involved in brain
biophysics~\cite{betzel2017multi, kandel2000principles}. In space,
these processes occur over centimeters at the organ level to
sub-nanometers at the atomic level inside cells
(\Cref{fig:brain_scales}). In time, processes occuring over seconds
and over hours may influence each other.

Traditionally, computational modeling of brain tissue has focused on
electrophysiology. An important reason for this is the impressively
accurate model of a giant squid axon potential developed by Hodgkin
and Huxley in 1952~\cite{hodgkin1952quantitative, hausser2000hodgkin,
  einevoll2006mathematical}. Analogies between electrical circuits and
neural tissue have resulted in volume conductor theory models,
compartmental models and network models, where extra- and
intracellular potentials are calculated based on modeling neurons as
compartments in a volume conductor or as components of electrical
circuits. These models often account for the geometry implicitly, by
including homogenized parameters such as volume fractions and
tortuosity. Multi-compartmental modeling based on cable theory has
developed into the standard framework for mechanistic modeling of
neurons~\cite{halnes2024electric}. Over the last decades, there has
been increasing interest in models accounting for geometry and
morphology explicitly~\cite{agudelo2013computationally,
  ellingsrud2020finite}. This development has taken place in tandem
with modeling cardiac cells and tissue, another example of excitable
tissues~\cite{tveito2021modeling}.

While most computational modeling of cerebral tissue has focused on
neurons, which are the main signaling units of the nervous
system~\cite{kandel2000principles}, in recent years, glial cells have
received an increasing amount of
attention~\cite{de2019computational}. Glial cells constitute at least
half of human brain tissue and are now considered to play an important
role in cerebral dynamics. For example, the glia known as astrocytes
respond to neurotransmitters, partake in modulation and extracellular
metabolism of adenosine triphosphate (ATP), and are involved in
neuronal signaling pathways through receptor dynamics
\cite{Guttenplan2025GPCRsignaling, Lefton2025NorepinephrineSignals,
  Chen2025NorepinephrineChanges}. Astrocytes also regulate brain
function through ion concentrations with examples including potassium
buffering~\cite{chen2000spatial} and the dynamic role of astrocytic
chloride concentrations in volume regulation and cell
proliferation~\cite{untiet2024astrocytic}. The ion concentrations both
in the extracellular and intracellular spaces of the brain play an
indispensable role in brain functioning, with up to two orders of
magnitude differences between physiological and pathological
concentrations~\cite{Rasmussen2020Interstitial}.

An open question relates to the role of electrodiffusion in brain
tissue under physiological and pathological conditions and the coupled
response of electric potentials and ion concentrations. Many
electrodiffusion models homogenize or idealize the brain tissue, often
considering zero-dimensional cellular compartments or rectangular or
cylindrical geometries~\cite{Halnes2013Electrodiffusive,
  solbraa2018kirchhoff, saetra2020electrodiffusive,
  saetra2021electrodiffusive, saetra2023neural,
  saetra2024electrodiffusive}. These models are particularly suitable
for studying the effect of ionic fluxes. For example, Halnes~\emph{et
al.}~\cite{Halnes2013Electrodiffusive} studied a one-dimensional model
of an astrocyte coupled with the extracellular space, stimulating the
system by adding potassium and removing sodium ions in the
astrocyte. Sætra~\emph{et al.}~\cite{saetra2021electrodiffusive} used
a similar stimulus for an electrodiffusive compartmental model with
neuronal, glial and extracellular spaces. There has also been
development of electrodiffusive models that explicitly represent the
cellular geometries~\cite{mori2009numerical, pods2013electrodiffusion,
  lu2010poisson, ellingsrud2020finite, lopreore2008computational}.
These partial differential equation (PDE)-based models are typically
solved numerically using finite difference, finite element, or finite
volume methods. The explicit geometry representation of individual
cells and membranes allows for studying the interplay between cellular
shape and function, a key tenet of cellular biology, and for studying
spatially varying membrane mechanisms such as distributed ion
channels, pumps or cotransporters.

Cell-by-cell models of ionic electrodiffusion are highly demanding to
construct and solve computationally~\cite{tveito2017evaluation}, for
several reasons. First, the morphology of neurons, astrocytes and the
extracellular space is highly complex. This makes generation of
well-defined, conforming computational meshes challenging in its own
right. Moreover, the fine geometric detail calls for high resolution
meshes, and the large number of mesh vertices and cells in turn
increases computational cost. Second, electrodiffusion involves
systems of PDEs with strongly coupled variables, whose dynamics evolve
on multiple spatial and temporal scales. Development of efficient and
scalable numerical solution algorithms for such coupled problems is an
active field of research~\cite{roy2023scalable, Benedusi2024Scalable,
  ellingsrud2025splitting}.

Over the last three decades, advances in bioimaging have
revolutionized our ability to study the structure and function of
living systems. At the nano- to microscale, the new capabilities of
cellular microscopy have resulted in the generation and wide
availability of large-scale high-resolution imaging data, as
strikingly exemplified by increasingly complex dense reconstructions
of cortical tissue. Through open distribution of such high-quality
imaging data, there now exists several datasets with extremely
detailed image data of brain tissue~\cite{microns2025functional}.
Together with modern image-processing software and powerful hardware,
the image data enables generation of realistic computational
geometries of brain tissue. These convergent developments give new
opportunities for detailed simulations of spatial ion dynamics in
brain tissue.

In this chapter, we present a computational framework for numerically
solving ionic electrodiffusion PDEs defined over extremely detailed
geometries of mouse visual cortex tissue. We highlight different
aspects of the computational pipeline: from the geometric properties
of these brain tissue to the numerical properties of the resulting
finite element discretization.

\section{Dense reconstructions of brain tissue from cell imaging}
\label{sec:reconstruction}

Brain imaging has been fundamental to our understanding of brain
structure and function~\cite{yen2023exploring}. Today, the combination
of highly detailed cell imaging data and computational processing
power allows for a new level of detail in terms of digital
representations of brain tissue. Neurons, glial cells and the
extracellular space are however characterized by tortuous, meandering
structures making meshing of such high-resolution digital
reconstructions of brain tissue challenging. One aspect is the
computational demand in processing the images, generating surfaces and
volumetric meshes. Another key element is that the generated
computational mesh must be of sufficient quality for numerical PDE
solution algorithms. We here illustrate how these challenges can be
addressed by current meshing technology by presenting a pipeline
generating conforming tetrahedral meshes from dense image
reconstructions.

\subsection{From segmented cell imaging to conforming finite element meshes}
\begin{figure}
    \centering
    \includegraphics[width=\textwidth]{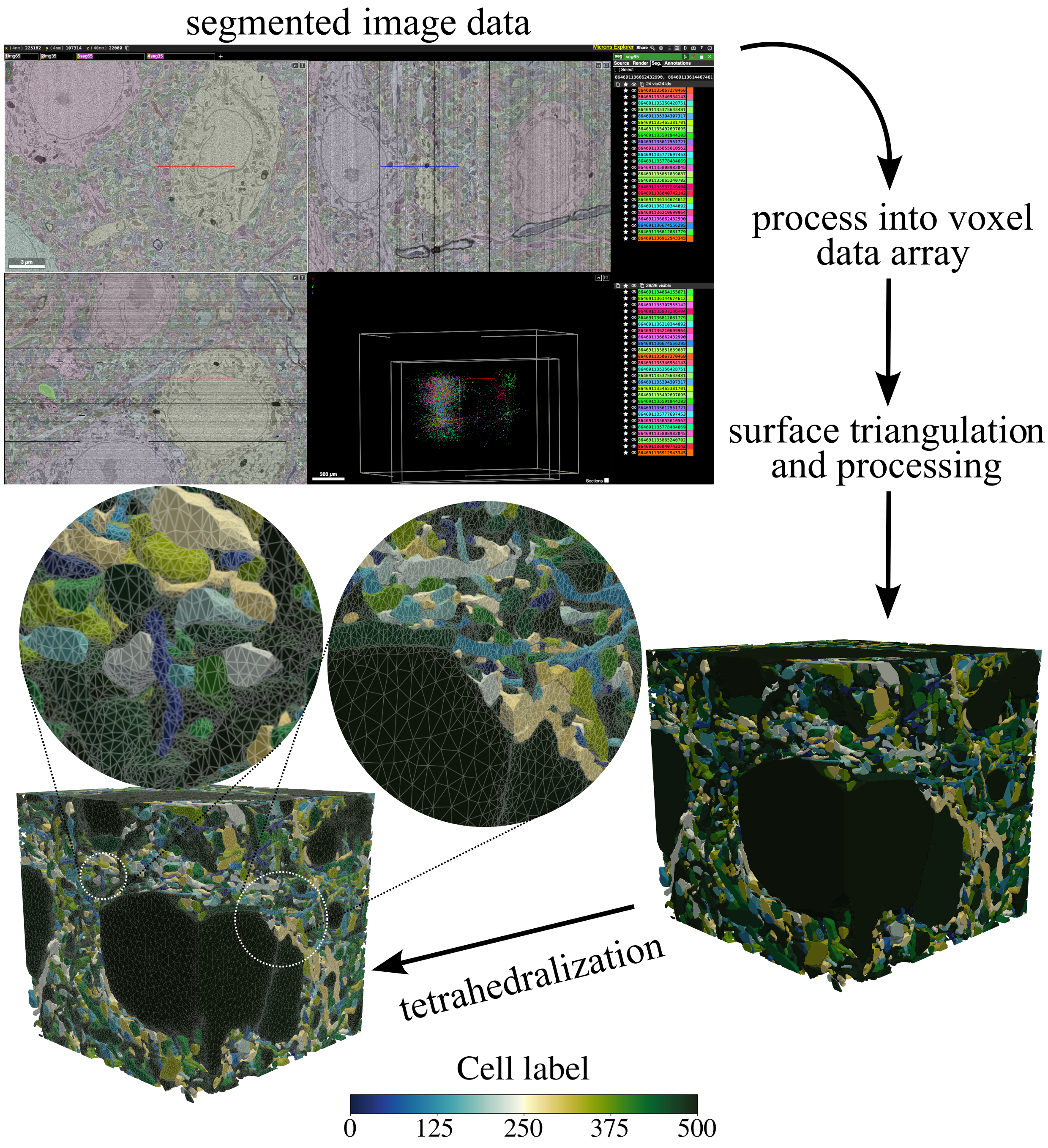}
    \caption{Illustration of the computational meshing pipeline.
      Segmented electron microscopy image data, here shown in the
      MICrONS explorer interface~\cite{microns2025functional}, is
      downloaded and processed into a voxel data array. The voxel
      data is next converted into triangulated surfaces representing
      the interfaces between the biological cells and extracellular
      space. These surfaces go through a sequence of morphological
      processing steps to ensure sufficient mesh quality
      downstream. Finally, a tetrahedral mesh is generated conforming
      to the interface surfaces. }
    \label{fig:seg_to_mesh}
\end{figure}

The pipeline is implemented in the open-source software
EMI-Meshing~\cite{Causemann2024EMIMeshing} and consists of the
following steps (see also~\Cref{fig:seg_to_mesh}):
\begin{enumerate}
    \item Select and extract segmented image data
    \item Filter and/or process the segmentation
    \item Construct triangulated surfaces representing the interface
      between intracellular and extracellular spaces
    \item Generate a tetrahedral mesh of the complete volume
      conforming to the interface surfaces.
\end{enumerate}
First, the raw segmentation data is downloaded at a voxel resolution
of $32\times 32 \times 40\,$nm. This data is subsequently resampled
to an isotropic resolution of $20\,$nm. To reduce noise and focus on
significant biological structures, we remove small cellular fragments
with volumes below 5000 voxels (approximately 0.04 $\mu$m$^3$) and
retain only the $N$ largest biological cells.

We then apply a sequence of morphological operations to process the
cellular shapes, utilizing the GPU-accelerated library
\emph{pyclesperanto}~\cite{Haase2025clEsperanto}. All operations are
performed with a radius of 1. The cells are first dilated into the
background, followed by morphological opening and closing operations
which effectively smooth the cellular boundaries. Crucially, all cells
are then eroded; this step ensures a physiological volume fraction for
the extracellular space and prevents contact between adjacent
cells. Following these morphological operations, any remaining
disconnected components smaller than the 5000 voxel threshold are
removed.

Surface meshes are extracted from the processed voxel data using the
marching cubes algorithm implemented in
\emph{PyVista}~\cite{sullivan2019pyvista}. To ensure computational
tractability, the resolution of these surface meshes is reduced by a
factor of ten using surface remeshing via the \emph{pyacvd}
library~\cite{Kaszynski2025pyacvd}. We mitigate staircase artifacts
arising from the voxel approximation by applying five iterations of
Taubin smoothing in \emph{PyVista}. Next, a cubic surface of edge
length L is generated to define the outer boundary of the
extracellular space.

Finally, we generate a volumetric mesh of the complete domain,
conforming to the triangulated interface surfaces to accurately
represent both the intracellular regions and the extracellular
space. This step is performed using the fTetWild
software~\cite{Hu2020FastWild} on the eX3 high-performance computing
infrastructure~\cite{ex3}. The resulting mesh resolution is governed
by the \emph{envelope size} and \emph{ideal edge length}
parameters. The \emph{envelope size} represents the maximal deviation
from the input surface, while the \emph{ideal edge length} defines the
target size of the generated tetrahedra. Using a small \emph{envelope
size} of 18 nm and a larger \emph{ideal edge length} (1--1.5 \% of the
bounding box length), the mesh resolution naturally adapts to the
shape and morphology of the cellular regions, such that larger
intracellular volumes are covered by larger mesh cells than the narrow
extracellular space gaps. To ensure element quality suitable for
numerical simulation, the mesh is optimized until a maximal conformal
AMIPS energy~\cite{fu2015computing} of 10 is reached, thereby
eliminating severely distorted tetrahedra. The segmentation labels are
transferred from the image data to the volumetric mesh, ensuring that
all biological cells and interface surfaces are uniquely labeled in
the final computational mesh. These labels or mesh tags facilitate
modelling different properties, membrane mechanisms, and initial and
boundary conditions for the different biological cells.

\begin{figure}
    \centering
    \includegraphics[width=\textwidth]{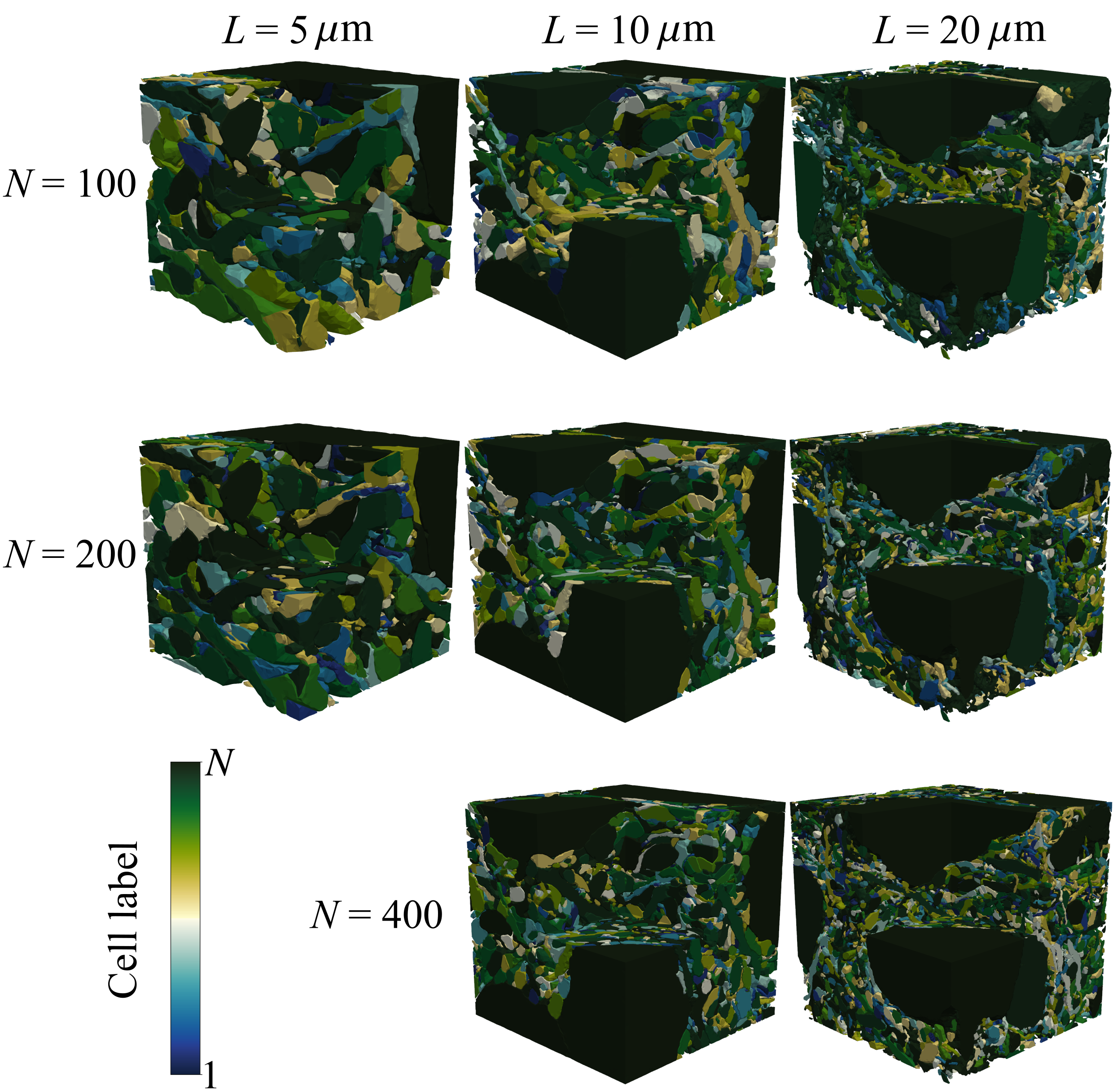}
    \caption{A sample of the computational meshes generated using the
      EMI-Meshing pipeline~\cite{Causemann2024EMIMeshing} of size $L
      \times L \times L$ containing $N$ biological cells surrounded by
      extracellular space.}
    \label{fig:geometries}
\end{figure}

We next apply this meshing pipeline to generate computational meshes
from the \textit{IARPA MICrONS Cortical $\text{mm}^3$} dataset
published by the MICrONS consortium~\cite{microns2025functional}.
This open dataset includes 3D images spanning a $1.4$mm $\times$
$0.87$mm $\times$ $0.84$mm volume of mouse visual cortex obtained
using two-photon microscopy, microCT and serial electron microscopy
(\Cref{fig:seg_to_mesh}). The image data are already segmented,
i.e.~each connected component within the 3D image has been identified
and assigned a unique label, and can be viewed using the
web-visualization tool Neuroglancer~\cite{neuroglancer}. We consider
a series of cubical volumes of reconstructed tissue of size $L \times
L \times L$ with center point $(225182, 107314, 22000)$ and with $L=5,
10, 20, 30$ $\mu$m. For each of these cubical volumes, we generate a
series of computational meshes including the $N$ largest biological
cells with $N=100, 200, 300, 400, 500$ (see
\Cref{fig:geometries}). The remaining volume is relabeled as
extracellular space.

\subsection{Most mesh vertices populate cellular membranes}

\begin{figure}
    \centering
    \includegraphics[width=\textwidth]{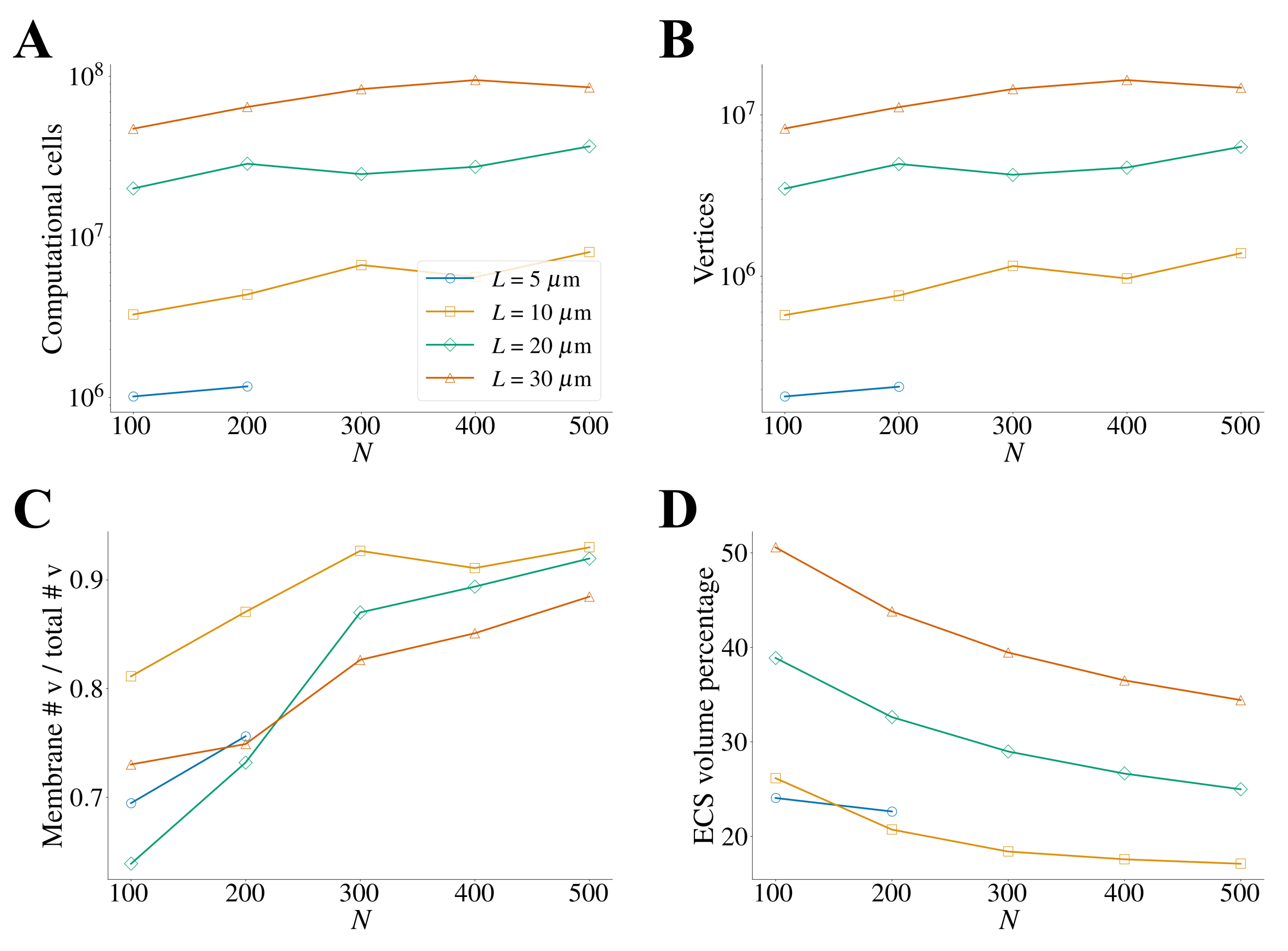}
    \caption{Statistics of cellular geometries of mouse visual cortex
      tissue. (\textbf{A}) Total number of computational cells in the
      mesh. (\textbf{B}) Total number of mesh vertices. (\textbf{C})
      The ratio of total number of membrane vertices to total number
      of mesh vertices. (\textbf{D}) Percentage of the total mesh
      volume that is occupied by extracellular space. Common legend
      for all subfigures. }
    \label{fig:mesh_statistics}
\end{figure}

We next analyze and present quantitative mesh characteristics for the
series of cubical meshes of the mouse visual cortex. When including an
increasing number of (biological) cells, the general trend for the
computational meshes is an increase in the number of computational
cells and vertices (\Cref{fig:mesh_statistics}A, B). However, the
increase is not uniform, for instance the number of mesh cells and
vertices is lower for the case $N = 400$ than for $N = 300$ when $L =
10$. Notably, the ratio of vertices located on the cell membranes to
the total number of vertices generally increases as the meshes become
denser. For the largest and densest meshes over 90\% of the mesh
vertices are on cellular membranes (\Cref{fig:mesh_statistics}C).
Moreover, the volume occupied by the extracellular space is reduced as
more biological cells are packed into the geometries
(\Cref{fig:mesh_statistics}D). This also results in a narrower and
more tortuous extracellular space. The ECS fraction of the densest
meshes are between 17--25\%. In real brain tissue, the fraction of
extracellular space is commonly reported around
20\%~\cite{chen2000spatial}. Note that for the mesh with $L=5$ $\mu$m,
we only include the $N=100$ and $N=200$ meshes, as the maximum number
of cells in this cube of tissue with the processing parameters we
employ is just shy of $200$ cells. For the mesh with size $L=10$
$\mu$m, we further present detailed statistics including the number of
ICS and ECS cells, and maximum and minimum edge lengths
(\Cref{tab:mesh_statistics}).

\begin{table}
     \caption{Statistics for the meshes with size $L=10$ $\mu$m
       containing $N$ biological cells ($N = 100, 200, 300, 400,
       500$. ICS: intracellular space, ECS: extracellular space.}
    \label{tab:mesh_statistics}
    \centering
    \begin{tabular}{l|rrrrr}
    \toprule
Quantity & $N = 100$ & $200$ & $300$ & $400$ & $500$ \\
    \midrule
    Number of cells             & 3 292 043 & 4 489 915 & 6 693 005 & 5 625 770 & 8 061 336 \\
    Number of cells in the ICS                       & 1 560 517 & 2 023 846 & 2 960 339 & 2 478 330 & 3 523 077 \\
    Number of cells in the ECS                       & 1 731 526 & 2 364 127 & 3 732 666 & 3 147 440 & 4 538 259 \\
    ECS volume fraction [\%]        &    26.2 &    20.7 &    18.4 &    17.6 &    17.1 \\
    Vertices                        &  574 314 &  759 501 & 1 158 188 &  968 230 & 1 389 747 \\
    Membrane vertices               &  465 906 &  661 225 & 1 073 278 &  881 889 & 1 292 304 \\
    Membrane vertices fraction [\%] &    81.1 &    87.1 &    92.7 &    91.1 &    93.0 \\
    Max edge length [nm]            &   456.1 &   435.6 &   449.8 &   422.6 &   450.9 \\
    Min edge length [nm]            &     3.6 &     2.3 &     5.3 &     4.5 &     3.9 \\
    \bottomrule
    \end{tabular}
\end{table}

\section{Modeling electrodiffusion at the cellular scale in explicit geometries}
With computational cellular geometries available, we next turn to
mathematical modeling of electrodiffusion. In a domain
$\Omega\subset\mathbb{R}^3$ representing cerebral tissue, we model ion
concentrations $c_k=c_k(\xx, t)$ with $k\in K$ for a set of ions $K =
\{\mathrm{Na}^+, \mathrm{K}^+, \mathrm{Cl}^{-}\}$ and the electric
potential $\phi=\phi(\xx, t)$, for $\xx\in\Omega$ and times $t\in[0,
  T]$. We will for notational brevity drop the valence of the ions in
the subscripts $k$. At the cellular scale, brain tissue is subdivided
into intracellular space, extracellular space, and the cellular
membranes that separate these spaces
(\Cref{fig:model_scales})~\cite{aidley1998physiology}. We first state
the general governing equations of electrodiffusion in cerebral
tissue, before extending the equations to intracellular and
extracellular spaces.

\subsection{Governing equations of electrodiffusion}
In the absence of fluid convection, the movement of ions
in brain tissue is described by the Nernst-Planck
equation, which states that the total flux $\JJ_k=\JJ_k(\xx, t)$
of an ion $k$ is a combination of
molecular diffusion and electrical drift~\cite{sterratt2023principles}:
\begin{equation}
    \JJ_k = \JJ_{k, \mathrm{diffusion}} + \JJ_{k, \mathrm{drift}} =
    -D_k\nabla c_k - \frac{z_k D_k F}{RT} c_k \nabla\phi.
    \label{eq:nernst_planck}
\end{equation}
Here $D_k$ and $z_k$ are the free diffusion coefficient and valence of
ion $k$, while $R = 8.314 \ \mathrm{Jmol^{-1}K^{-1}}$ is the universal
gas constant, $T = 300 \ \mathrm{K}$ is absolute temperature and $F=96
485 \ \mathrm{Cmol^{-1}}$ is Faraday's constant. By assuming
conservation of ions in the domain $\Omega$, it follows that
\begin{equation}
    \pdifft{c_k} + \nabla\cdot\JJ_k = f_k \quad\text{in }\Omega,
    \label{eq:conservation_k}
\end{equation}
where $f_k = f_k(\xx, t)$ is a rate of addition or removal of ions. 

An equation for the electric potential is derived by
assuming bulk electroneutrality in the domain $\Omega$.
This is equivalent to no charge separation, meaning
that the total sum of ion charges equals zero:
\begin{equation}
    F\sum_{k=1}^K z_k c_k = 0.
    \label{eq:electroneutrality}
\end{equation}
One can derive an alternative form of \eqref{eq:electroneutrality} by
differentiating the sum with respect to time, invoking
\eqref{eq:conservation_k}, and assuming that the source terms $f_k$
are electroneutral. This gives
\begin{equation}
    F\sum_{k=1}^K z_k \nabla\cdot\JJ_k = 0.
    \label{eq:electroneutrality_elliptic}
\end{equation}
In a discrete setting, the ellipticity of \eqref{eq:electroneutrality_elliptic} 
can make this formulation favorable compared to \eqref{eq:electroneutrality}.
Coupling \eqref{eq:electroneutrality} or \eqref{eq:electroneutrality_elliptic}
with \eqref{eq:nernst_planck} and \eqref{eq:conservation_k} has been studied in
several works under several names~\cite{mori2009numerical, halnes2016effect,
roy2023scalable, ellingsrud2020finite, solbraa2018kirchhoff}. Here we refer to
this system of equations as the Kirchhoff-Nernst-Planck (KNP) equations.

\subsection{Governing equations in multiple cells}
\begin{figure}
    \begin{minipage}[b]{0.45\textwidth}
        \centering
        \begin{tikzpicture}[scale=1.0]

    \draw[draw=black,very thick,fill=blue!10!white] (0,0) rectangle (4,4);
    \draw[draw=black,very thick,fill=red!10!white] (1,1) rectangle (3,3);

    \node at (1.7,0.3) {$\Omega_e$};
    \node at (1.7,2.2) {$\Omega_i$};
    \node at (0.6,1.5) {$\Gamma$};
    \node at (-0.5,1) {$\partial\Omega$};

    \draw[] (0.8,1.5) -- (1,1.5);
    \draw[] (-0.2,1) -- (0,1);


    \draw[gray, very thin] (0,0) -- (4,4);
    \draw[gray, very thin] (1,0) -- (4,3);
    \draw[gray, very thin] (2,0) -- (4,2);
    \draw[gray, very thin] (3,0) -- (4,1);
    \draw[gray, very thin] (0,1) -- (3,4);
    \draw[gray, very thin] (0,2) -- (2,4);
    \draw[gray, very thin] (0,3) -- (1,4);

    \draw[gray, very thin] (1,0) -- (1,4);
    \draw[gray, very thin] (2,0) -- (2,4);
    \draw[gray, very thin] (3,0) -- (3,4);

    \draw[gray, very thin] (0,1) -- (4,1);
    \draw[gray, very thin] (0,2) -- (4,2);
    \draw[gray, very thin] (0,3) -- (4,3);

\end{tikzpicture}
    \end{minipage}
    \hfill 
    \begin{minipage}[b]{0.45\textwidth}
        \centering
        \begin{tikzpicture}[scale=1.0]

    \draw[draw=black,very thick] (0,0) rectangle (4,4);
    \draw[draw=black,very thick] (1,1) rectangle (3,3);

    \node at (1.8,0.4) {$V_{e}(\Omega_e)$};
    \node at (1.8,2.4) {$V_{i}(\Omega_i)$};



    \draw[gray, very thin] (0,0) -- (4,4);
    \draw[gray, very thin] (1,0) -- (4,3);
    \draw[gray, very thin] (2,0) -- (4,2);
    \draw[gray, very thin] (3,0) -- (4,1);
    \draw[gray, very thin] (0,1) -- (3,4);
    \draw[gray, very thin] (0,2) -- (2,4);
    \draw[gray, very thin] (0,3) -- (1,4);

    \draw[gray, very thin] (1,0) -- (1,4);
    \draw[gray, very thin] (2,0) -- (2,4);
    \draw[gray, very thin] (3,0) -- (3,4);
    \draw[gray, very thin] (0,1) -- (4,1);
    \draw[gray, very thin] (0,2) -- (4,2);
    \draw[gray, very thin] (0,3) -- (4,3);

    \filldraw [red] (0,0) circle (2pt);
    \filldraw [red] (0,1) circle (2pt);
    \filldraw [red] (0,2) circle (2pt);
    \filldraw [red] (0,3) circle (2pt);
    \filldraw [red] (0,4) circle (2pt);
    \filldraw [red] (4,0) circle (2pt);
    \filldraw [red] (4,1) circle (2pt);
    \filldraw [red] (4,2) circle (2pt);

    \filldraw [red] (4,3) circle (2pt);
    \filldraw [red] (4,4) circle (2pt);
    \filldraw [red] (1,0) circle (2pt);
    \filldraw [red] (2,0) circle (2pt);
    \filldraw [red] (3,0) circle (2pt);
    \filldraw [red] (1,4) circle (2pt);
    \filldraw [red] (2,4) circle (2pt);
    \filldraw [red] (3,4) circle (2pt);


    \filldraw [orange] (0.95,0.95) circle (2pt);
    \filldraw [orange] (0.95,2) circle (2pt);
    \filldraw [orange] (0.95,3.05) circle (2pt);
    \filldraw [orange] (3.05,0.95) circle (2pt);
    \filldraw [orange] (3.05,2) circle (2pt);
    \filldraw [orange] (3.05,3.05) circle (2pt);
    \filldraw [orange] (2,0.95) circle (2pt);
    \filldraw [orange] (2,3.05) circle (2pt);
    \filldraw [cyan] (2,2) circle (2pt);

    \filldraw [red] (4.7,3.1) circle (2pt);
    \filldraw [orange] (4.7,2.4) circle (2pt);
    \filldraw [cyan] (4.7,1.7) circle (2pt);

    \node at (5.2,2.75) {$N_e$};
    \node at (5.2,2.05) {$N_{\Gamma}$};
    \node at (5.2,1.35)   {$N_i$};

\end{tikzpicture}
    \end{minipage}
    \caption{Conceptual illustration of the geometry and meshes. Left:
      idealized geometry, illustrating the intracellular ($\Omega_i$)
      and extracellular ($\Omega_e$) spaces separated by a membrane
      $\Gamma$.  The boundary of the domain is $\partial\Omega$.
      Right: The degree of freedom layout of the idealized cellular
      geometry when discretized with linear continuous Lagrange
      elements.  The function spaces $V_i(\Omega_i)$ and
      $V_e(\Omega_e)$ share the degrees of freedom on $\Gamma$.}
    \label{fig:model_scales}
\end{figure}
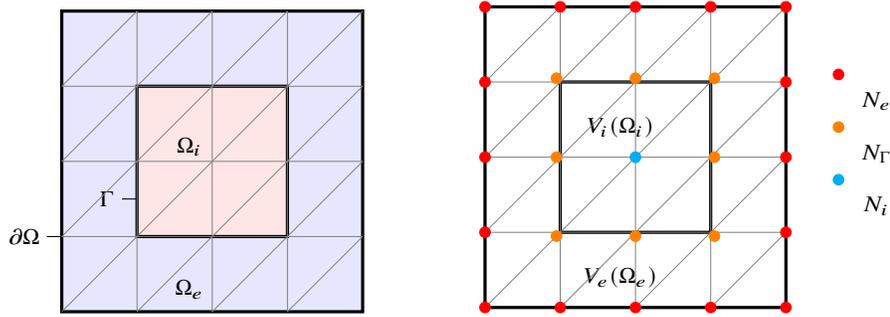

\Cref{eq:nernst_planck,eq:conservation_k,eq:electroneutrality_elliptic}
describe electroneutral electrodiffusion in the bulk of a domain
$\Omega$. We now turn to modeling excitable tissue where multiple
cells communicate through the extracellular space across the cellular
membranes. The model is based on the
Extracellular-Membrane-Intracellular (EMI) framework, which considers
separate domains for the extracellular space, the intracellular
spaces, and the cellular membranes~\cite{tveito2021modeling}. In
principle, there can be $N$ intracellular domains $\Omega_{i, j}$ for
$j=1, 2 \dots, N$ representing biological cells with a common
extracellular domain $\Omega_e$, with cellular membranes $\Gamma_j$
denoting the interface between $\Omega_{i, j}$ and $\Omega_e$.  In
this section, we describe the case with one biological cell ($N=1$,
\Cref{fig:model_scales}). The generalization to $N$ non-intersecting
biological cells is immediate.

Let $c_{k, r} = c_{k, r}(\xx, t)$ be the ion concentrations and
$\phi_r=\phi_r(\xx, t)$ the electric potential in a domain $\Omega_r$,
where $r=\{i, e\}$ denotes the intra- or extracellular space. We
continue to consider the ion set $K = \{\mathrm{Na}^+, \mathrm{K}^+,
\mathrm{Cl}^-\}$. \Cref{eq:nernst_planck,eq:conservation_k,eq:electroneutrality_elliptic}
describe electrodiffusion in the bulk of the domains $\Omega_r$:
\begin{alignat}{2}
    \pdifft{c_{k, r}} + \nabla\cdot\JJ_{k, r} &= f_{k, r} \quad&\text{in }\Omega_r,
    \label{eq:conservation_k_in_r} \\
    F\sum_{k=1}^K z_k \nabla\cdot\JJ_{k, r} &= 0 \quad&\text{in }\Omega_r,
    \label{eq:electroneutrality_in_r}
\end{alignat}
with the flux $\JJ_{k, r}$ given by the Nernst-Planck equation
\eqref{eq:nernst_planck}. As we have coupled the KNP equations with
the EMI framework, we refer to \eqref{eq:conservation_k_in_r} and
\eqref{eq:electroneutrality_in_r} as the KNP-EMI equations.

The intra- and extracellular dynamics are coupled at the cellular
membrane $\Gamma$ that separates the intra- and extracellular spaces.
Cellular membranes in cerebral tissue are not freely permeable to
ions, but lined with selective ion channels that only permit transport
of select ions under specific circumstances. Because of this, there
are spatial differences in the intra- and extracellular ion
concentrations and thus a charge difference across the membrane. The
concentration differences induce an electric potential difference
across the membrane, labeled the membrane potential $\phi_m = \phi_i -
\phi_e$, and the membrane acts like a capacitor storing electric
energy~\cite{sterratt2023principles}. On the cellular membrane
$\Gamma$, we assume continuity in the membrane current density $I_m$:
\begin{equation}
    I_m =  F\sum_{k\in K} z_k\JJ_{k, i}\cdot\nn_i
        = -F\sum_{k\in K} z_k\JJ_{k, e}\cdot\nn_e, 
        \quad\text{on }\Gamma
    \label{eq:KNP_membrane_current}
\end{equation}
where the normal vector $\nn_r = \nn_r(\xx)$ is directed out of the
domain $\Omega_r$.

We will here consider the \emph{single-dimensional} form of the
KNP-EMI equations, where the membrane current density $I_m$ that
exists on a lower-dimensional manifold is eliminated from the system
of equations \cite{tveito2021modeling, Benedusi2024Scalable,
  ellingsrud2020finite}.  We assume that the membrane current $I_m$ is
the sum of an ionic current $I_{\mathrm{ion}}$, whose model depends on
the physiological scenario considered, and a capacitive current
$I_{\mathrm{cap}}$, which reflects the electric energy stored in the
membrane.  Assuming that the membrane is a parallel plate capacitor
with capacitance $C_m$, the capacitive current is
\begin{equation}
    I_{\mathrm{cap}} = C_m\pdifft{\phi_m}.
    \label{eq:EMI_capacitive_current}
\end{equation}
We assume that both the ionic current and the capacitive current
has ion-specific contributions:
\begin{equation}
    I_{\mathrm{ion}} + I_{\mathrm{cap}} = \sum_{k\in K}I_{\mathrm{ion}, k} + \sum_{k\in K} I_{\mathrm{cap}, k},
    \label{eq:KNP_current_ion_split}
\end{equation}
with 
\begin{equation}
    I_{\mathrm{cap}, k} = \alpha_{k, r}I_{\mathrm{cap}}
    \label{eq:capacitive_current_k}
\end{equation}
weighted by the coefficients
\begin{equation}
    \alpha_{k, r} = \frac{z_k^2D_{k, r}c_{k, r}}{\sum_{l\in K}z_l^2 D_{l, r}c_{l, r}}.
    \label{eq:alpha_ratio}
\end{equation}
By definition $\sum_{k\in K}\alpha_{k, r} = 1$. Using
\eqref{eq:alpha_ratio} and \eqref{eq:capacitive_current_k} with the
relation \eqref{eq:KNP_current_ion_split} and the definition of the
membrane current \eqref{eq:KNP_membrane_current}, the flux of an ion
$k$ is expressed as a function of the ionic currents and the membrane
potential
\begin{equation}
    \JJ_{k, r}\cdot\nn_r = \pm_r\frac{I_{\mathrm{ion}, k} + I_{\mathrm{cap}, k}}{F z_k}
    = \pm_r\frac{I_{\mathrm{ion}, k} + \alpha_{k, r}C_m\pdifft{\phi_m}}{F z_k}
\end{equation}
for $r\in\{i, e\}$, with the signs $\pm_i = 1$ and $\pm_e = -1$. 

\subsection{Membrane mechanisms and stimulus}
\label{subsec:membrane_mechanisms_and_stimuli}

Cellular membrane mechanisms such as ion channels, pumps and
cotransporters are central to excitable tissue dynamics.  The membrane
mechanisms determine how cells respond to external stimuli such as
synaptic signaling, or to changes in the extracellular environment.
When modelling membrane mechanisms, one considers the cell types and
biological scenarios at hand. For example, neurons and astrocytes have
different membrane mechanisms~\cite{aidley1998physiology,
  Halnes2013Electrodiffusive}. Here, we let neuronal membrane dynamics
be governed by the Hodgkin-Huxley
model~\cite{hodgkin1952quantitative}, an
$\mathrm{Na}^+$/$\mathrm{K}^+$-ATPase pump~\cite{kager2000simulated}
and the two cotransporters KCC2 and NKCC1~\cite{wei2014unification,
  ellingsrud2020cell}.  Glial membrane dynamics are governed by an ATP
pump~\cite{Halnes2013Electrodiffusive}, passive leak channels with an
inward-rectifying potassium current~\cite{Halnes2013Electrodiffusive},
and the KCC1 and NKCC1 cotransporters~\cite{ostby2009astrocytic}.

The Hodgkin-Huxley model~\cite{hodgkin1952quantitative} enables
simulating action potentials, the communicating signal of neurons.
The model includes ion channel currents that are voltage-gated;
i.e.,~whether they are open or closed depend on the difference between
a resting membrane potential $\phi_{\mathrm{rest}}$ and an
ion-specific reversal potential $E_k$. The Hodgkin-Huxley model
includes passive (leak) channels and active channels.  Letting $g_{k,
  \mathrm{leak}}$ be the conductance of the leak channel of ion $k$,
the general expression for a neuronal leak channel current is
\begin{equation*}
    I_{\mathrm{leak}, k, n} = g_{\mathrm{leak}, k, n}(\phi_m - E_k).
\end{equation*}
The reversal potential
\begin{equation*}
    E_k = \frac{RT}{z_k F}\log{\frac{c_{k, e}}{c_{k,i}}},
\end{equation*}
follows from the Nernst-Planck equation, considering a steady state
where the diffusive flux balances the electrical drift across the
cellular membrane~\cite{sterratt2023principles}.

The active channels include gating variables $n, m$ and $h$ that reflect the
probability that an ion channel is open or closed. For sodium and potassium,
the active channel currents are
\begin{align*}
    I_{\mathrm{active, Na}} &= \overline{g}_{\mathrm{Na}}m^3h(\phi_m - E_{\mathrm{Na}}), \\
    I_{\mathrm{active, K}}  &= \overline{g}_{\mathrm{K}} n^4 (\phi_m - E_{\mathrm{K}}) ,
\end{align*}
where $\overline{g}_k$ is the maximum conductivity of the channel for
ion $k$.  There is no active channel for chloride.  Time evolution of
the gating variables $n, m, h$ is governed by ordinary differential
equations (ODEs) of the general form
\begin{equation}
    \pdifft{\zeta} = \alpha_{\zeta}(1 - \zeta) - \beta_{\zeta}\zeta,
    \label{eq:gating_variable_ODE}
\end{equation}
where $\zeta = (n(t), m(t), h(t))$. The coefficients $\alpha_{\zeta}$
and $\beta_{\zeta}$ depend on the difference $\phi_m -
\phi_{\mathrm{resting}}$. The ODEs \eqref{eq:gating_variable_ODE} are
equipped with appropriate initial conditions $\zeta(t=0) = \zeta_0 =
(n_0, m_0, h_0)$. See \cite{hodgkin1952quantitative} for full
comprehensive details of the model and its coefficients.

In addition to passive and active ion channels, which are driven by
the difference $\phi_m - E_k$, we model membrane mechanisms that are
driven by differences in intra- and extracellular concentrations.
First, we consider the sodium-potassium pump
$\mathrm{Na}^+$/$\mathrm{K}^+$-ATPase.  This pump uses ATP (adenosine
triphosphate) as energy to pump sodium out of a cell and potassium
into the cell. For every ATP molecule consumed, three sodium ions are
exported and two potassium ions are imported, meaning the pump is a
net exporter of electric charge. The membrane current density
associated with the ATP pump on neuronal cell membranes is modeled by
\begin{equation*}
    I_{\mathrm{ATP}, n} = \frac{S_{\mathrm{ATP}, n}}{(1 + P_{\mathrm{K}}/c_{{\mathrm{K}, e}})^2
    (1 + P_{\mathrm{Na}}/c_{\mathrm{Na}, i})^3},
\end{equation*}
where $S_{\mathrm{ATP}, n} = 0.25 \ \mathrm{A{m}^{-2}}$ is the pump
strength, while $P_{\mathrm{K}} = 1.5$ mM and $P_{\mathrm{Na}}=10$ mM
are threshold values that determine when the pump strength increases
and decreases~\cite{oyehaug2012dependence,kager2000simulated}.

For the neuronal membranes, we also consider the two cotransporters
NKCC1 and KCC2.  The NKCC1 cotransporter transports sodium, potassium
and chloride into the cell, while the KCC2 cotransporter transports
potassium and chloride out of cells~\cite{wei2014unification,
  ellingsrud2020cell, oyehaug2012dependence}.  The membrane current
densities associated with the cotransporters are defined as
\begin{align*}
    I_{\mathrm{NKCC1}, n} &= \delta S_{\mathrm{NKCC1}, n}
    \ln{\frac{c_{\mathrm{K}, e}c_{\mathrm{Cl}, e}}{c_{\mathrm{K}, i}c_{\mathrm{Cl}, i}}}, \\
    I_{\mathrm{KCC2}} &= S_{\mathrm{KCC2}}
    \ln{\frac{c_{\mathrm{Na}, e}c_{\mathrm{K}, e}c_{\mathrm{Cl}, e}^2}{c_{\mathrm{Na}, i}c_{\mathrm{K}, i}c_{\mathrm{Cl}, i}^2}},
\end{align*}
where $S_{\mathrm{NKCC1}, n} = 0.0023 \ \mathrm{A{m}^{-2}}$ and
$S_{\mathrm{KCC2}} = 0.0068 \ \mathrm{A{m}^{-2}}$ are the maximum
strengths of the cotransporters. The function $\delta =
\delta(c_{\mathrm{K}, e}^0, c_{\mathrm{K}, e})$ is defined as
\begin{equation}
    \delta(c_{\mathrm{K}, e}^0, c_{\mathrm{K}, e}) = \frac{1.0}{(1.0 + (0.03 / (c_{\mathrm{K}, e} - c_{\mathrm{K}, e}^0))^{10})}
    \label{eq:nkcc1_delta}
\end{equation}
if $c_{\mathrm{K}, e}\in[3.0, c_{\mathrm{K}, e}^0]$, else $\delta = 0$
to suppress activity of the NKCC1
cotransporter~\cite{oyehaug2012dependence}.

For glial cells, we consider mechanisms described in previous
astrocyte modeling studies \cite{ostby2009astrocytic,
  Halnes2013Electrodiffusive, oyehaug2012dependence}.  We define the
glial leak channel currents with different conductances than for the
neuronal cells
\begin{equation*}
    I_{\mathrm{leak}, k, g} = g_{\mathrm{leak}, k, g}f_{\mathrm{Kir-Na}, k}(\phi_m - E_k),
\end{equation*}
where
\begin{align}
    f_{\mathrm{Kir-Na}, k} = 
    \begin{cases}
        1.0, &\quad\text{for }k\in\{\text{Na, Cl}\}, \\
        \sqrt{\frac{c_{\mathrm{K}, e}}{c_{\mathrm{K}, e}^0}}\frac{AB}{CD}, &\quad\text{for }k = \mathrm{K}.
    \end{cases}
    \label{eq:f_kir_na}
\end{align}
This function is used to model the potassium leak channel as
inward-rectifying~\cite{Halnes2013Electrodiffusive}.
In~\Cref{eq:f_kir_na}, $c_{\mathrm{K}, e}^0$ is the initial
extracellular potassium concentration, while the coefficients $A, B,
C$ and $D$ are defined as
\begin{align*}
    A &= 1 + e^{0.433}, \\
    B &= 1 + e^{\frac{-(0.1186 + E_{\mathrm{K}}^0)}{0.0441}}, \\
    C &= 1 + e^{\frac{(\phi_m - E_{\mathrm{K}} + 0.0185)}{0.0425}}, \\
    D &= 1 + e^{\frac{-(0.1186 + \phi_m)}{0.0441}},
\end{align*}
where $E_{\mathrm{K}}^0$ is the initial potassium Nernst potential.

We model a $\mathrm{Na}^+$/$\mathrm{K}^+$-ATPase pump current and
cotransporters for glia, similar to the approach for neurons.  The
glial ATP pump current density is expressed as
\begin{equation*}
    I_{\mathrm{ATP}, g} = \frac{S_{\mathrm{ATP}, g}}
    {(1 + P_{\mathrm{K}}/c_{\mathrm{K}, e})
    (1 + P_{\mathrm{Na}}/c_{\mathrm{Na}, i})^{3/2}},
\end{equation*}
where $S_{\mathrm{ATP}, g} = 0.12 \ \mathrm{A m^{-2}}$,
$P_{\mathrm{Na}} = 10$ mM and $P_{\mathrm{K}} = 1.5$ mM are the
maximum pump strength, sodium and potassium pump thresholds,
respectively~\cite{ostby2009astrocytic, Halnes2013Electrodiffusive}.

For glia, we consider the NKCC1 and KCC1
cotransporters~\cite{ostby2009astrocytic, oyehaug2012dependence}.  The
membrane current densities associated with these cotransporters are
\begin{align*}
    I_{\mathrm{NKCC1}, g} &= \delta S_{\mathrm{NKCC1}, g}
    \ln{\frac{c_{\mathrm{K}, e}c_{\mathrm{Cl}, e}}{c_{\mathrm{K}, i}c_{\mathrm{Cl}, i}}}, \\
    I_{\mathrm{KCC1}} &= S_{\mathrm{KCC1}}
    \ln{\frac{c_{\mathrm{Na}, e}c_{\mathrm{K}, e}c_{\mathrm{Cl}, e}^2}{c_{\mathrm{Na}, i}c_{\mathrm{K}, i}c_{\mathrm{Cl}, i}^2}},
\end{align*}
where the maximum cotransporter strengths are $S_{\mathrm{NKCC1}, g} =
\num{5.2e-4} \ \mathrm{Am^{-2}}$ and $S_{\mathrm{KCC1}} = 0.018
\ \mathrm{Am^{-2}}$, and $\delta$ is given by \Cref{eq:nkcc1_delta}.

We stimulate the brain tissue through adding a stimulus current
density to the sodium ionic current density. We model an exponentially
decreasing stimulus with the expression
\begin{equation}
    I_{\mathrm{stim}} = \overline{g}_{\mathrm{stim}} e^{(-t \ \mathrm{mod} \ T_{\mathrm{stim}}) / a_{\mathrm{stim}}}.
    \label{eq:stimulus_current_density}
\end{equation}
Here $\overline{g}_{\mathrm{stim}}$ is the maximum stimulus
conductance.  The exponential factor
in~\Cref{eq:stimulus_current_density} makes $I_{\mathrm{stim}}$ decay
exponentially with a time constant $a_{\mathrm{stim}} = 0.5$ ms, and
the modulus operation is used to turn the current on periodically with
a period $T_{\mathrm{stim}}$.

To summarize, the ionic current densities for neuronal cell membranes are
\begin{align*}
    I_{\mathrm{ion, Na}, n} &= I_{\mathrm{leak, Na}, n} + I_{\mathrm{active, Na}}
    + I_{\mathrm{stim}} + 3I_{\mathrm{ATP}, n} - I_{\mathrm{NKCC1}, n}, \\
    I_{\mathrm{ion, K}, n}  &= I_{\mathrm{leak, K}} + I_{\mathrm{active, K}}
    - 2I_{\mathrm{ATP}, n} - I_{\mathrm{NKCC1}, n} - I_{\mathrm{KCC2}}, \\
    I_{\mathrm{ion, Cl}, n} &= I_{\mathrm{leak, Cl}} + 2I_{\mathrm{NKCC1}, n} + I_{\mathrm{KCC2}},
\end{align*}
while in the cases where we consider glial cells, the glial ionic
membrane current densities are
\begin{align*}
    I_{\mathrm{ion, Na}, g} &= I_{\mathrm{leak, Na}, g} + I_{\mathrm{active, Na}}
    + 3I_{\mathrm{ATP}, g} - I_{\mathrm{NKCC1}, g}, \\
    I_{\mathrm{ion, K}, g}  &= I_{\mathrm{leak, K}}
    - 2I_{\mathrm{ATP}, g} - I_{\mathrm{NKCC1}, g} - I_{\mathrm{KCC1}}, \\
    I_{\mathrm{ion, Cl}, g} &= I_{\mathrm{leak, Cl}, g} + 2I_{\mathrm{NKCC1}, g} + I_{\mathrm{KCC1}}.
\end{align*}

\subsection{Boundary conditions}
With the membrane currents defined, it remains to determine boundary
and initial conditions to close the KNP-EMI equation system. As
boundary conditions, we consider a no-flux condition on all boundaries
$\partial\Omega_r$ exterior to the domain $\Omega_r$ for all times $t
> 0$:
\begin{equation}
    \JJ_{k, r} \cdot \nn_r = 0, \quad\forall \, k \quad\text{on }\partial\Omega_r, \quad\forall r.
    \label{eq:no_flux_BC}
\end{equation}
With the boundary condition \eqref{eq:no_flux_BC} enforced on all
boundaries, the electric potentials are only determined up to a
constant. We return to this point in \Cref{subsec:iterative_solver}.

\subsection{Initial conditions and defining resting states}
As initial conditions, we set the initial concentrations $c_{k,
  r}(\xx, 0) = c_{k, r}^0$ and electric potentials $\phi_r(\xx, 0) =
\phi_r^0$.

To define initial conditions that constitute a resting state with no
changes in ionic concentrations or electrical potentials, we solve a
system of ordinary differential equations (ODEs) that represent the
dynamics in a single spatial point.  When only considering neurons, we
collapse the intracellular spaces and the extracellular space into two
separate compartments.  When modeling both neuronal and glial cells,
we consider two separate intracellular compartments, one for each cell
type, and a third compartment for the extracellular space. In these
compartmental models, the change in the membrane potential at steady
state is given by the total ionic current $I_{\mathrm{ion}}$, since
then the total membrane current density $I_m = I_{\mathrm{ion}} +
I_{\mathrm{cap}} = 0$.  The ion concentration changes are in turn
given by the molar fluxes associated with the total ionic current
densities of the respective ions.  The resulting steady-state ODE
system for the three-compartment model is
\begin{equation}
    \begin{aligned}
        \frac{\mathrm d\phi_{m, l}}{\mathrm dt}  &= -\frac{1}{C_m} I_{\mathrm{ion}, l}, \\
        \frac{\mathrm d c_{k, i, l}}{\mathrm dt} &= -\frac{I_{\mathrm{ion}, k, l}}{z_k F}\frac{|\Gamma|_{l}}{|\Omega|_{i, l}}, \\
        \frac{\mathrm d c_{k, e}}{\mathrm dt}    &=  \frac{1}{z_k F}
                                                        \left(
                                                            I_{\mathrm{ion}, k, n}\frac{|\Gamma|_{n}}{|\Omega|_{e}} 
                                                          + I_{\mathrm{ion}, k, g}\frac{|\Gamma|_{g}}{|\Omega|_{e}}
                                                        \right), \\
        \frac{\mathrm d\zeta}{\mathrm dt}        &= \alpha_{\zeta}(1 - \zeta) - \beta_{\zeta}\zeta.
    \end{aligned}
    \label{eq:initial_condition_ODE_system}
\end{equation}
This ODE system respects our convention of considering currents out of
a cell as negative fluxes. The subscript $l\in\{n, g\}$ denotes
neuronal or glial cell, $|\Gamma|_{l}$ is the total surface area of
the cellular membranes separating the compartment of cell type $l$
from the extracellular space, $|\Omega|_{i, l}$ is the total volume of
all intracellular compartments of cell type $l$, and $|\Omega|_{e}$ is
the volume of the extracellular space.  The two-compartment system
results from reducing \eqref{eq:initial_condition_ODE_system} by
considering only $l=n$.  A steady-state solution to the system
\eqref{eq:initial_condition_ODE_system} will constitute a steady state
for the full PDE system, given that we set the initial condition
constant in space and that the membrane mechanisms are homogeneous in
space.

To determine initial conditions for the PDE system by solving the ODE
system \eqref{eq:initial_condition_ODE_system}, we consider a set of
initial conditions for the ODE system. The initial membrane potentials
used to solve the ODE system are $-70$ mV and $-85$ mV for the neurons
and astrocytes, respectively (taken from~\cite{Benedusi2024Scalable,
  oyehaug2012dependence, oyehaug2023slow}).  The initial
concentrations are based on physiological concentrations in neurons
and glial cells (taken from~\cite{neuroscienceBook, oyehaug2023slow,
  oyehaug2012dependence, ostby2009astrocytic}).  Finally, the initial
conditions for the gating variables are set by solving
\eqref{eq:gating_variable_ODE} at steady state with $\phi_m = -70$ mV.
All the membrane mechanism parameters and physical constants in the
KNP-EMI model used in this work are listed in~\Cref{tab:parameters}.
We note that some of the membrane mechanism parameters were calibrated
to achieve balance between the membrane mechanisms.
\begin{table}[!t]
    \caption{Material and model parameters. (*) Used to determine
    initial conditions with a steady-state approximation.
    (\textdagger) Calibrated based on reference.
    }
    \label{tab:parameters}
    \centering
    \begin{tabular}{p{5.9cm}p{1.4cm}p{1.7cm}p{1.5cm}p{1.1cm}}
    \toprule
        Parameter & Symbol & Value & Unit & Ref. \\
        \midrule
        Universal gas constant & $R$ & 8.314 & J$\mathrm{K^{-1}mol^{-1}}$ & \\
    Faraday constant       & $F$ & 96485 & C$\mathrm{mol}^{-1}$ & \\
    Temperature            & $T$ &   300 & K &   \\
    Free diffusion coefficient sodium & $D_{r, \mathrm{Na^+}}$ & \num{1.33e-9} & $\mathrm{m^2s^{-1}}$ & \cite{hille2001ion} \\
    Free diffusion coefficient potassium & $D_{r, \mathrm{K^+}}$  & \num{1.96e-9} & $\mathrm{m^2s^{-1}}$ & \cite{hille2001ion} \\
    Free diffusion coefficient chloride & $D_{r, \mathrm{Cl^-}}$ & \num{2.03e-9} & $\mathrm{m^2s^{-1}}$ & \cite{hille2001ion} \\
    Initial neuronal sodium concentration    (ICS) & $[\mathrm{Na^+}]_{i, n}^0$  &   10  & mM & \cite{neuroscienceBook}* \\
    Initial glial    sodium concentration    (ICS) & $[\mathrm{Na^+}]_{i, g}^0$  &   15  & mM & \cite{oyehaug2012dependence}* \\
    Initial          sodium concentration    (ECS) & $[\mathrm{Na^+}]_e^0$       &  145  & mM & \cite{oyehaug2012dependence}* \\
    Initial neuronal potassium concentration (ICS) & $[\mathrm{K^+}]_{i, n}^0$   &  130  & mM & \cite{neuroscienceBook}* \\
    Initial glial    potassium concentration (ICS) & $[\mathrm{K^+}]_{i, g}^0$   &  100  & mM & \cite{oyehaug2012dependence}* \\
    Initial          potassium concentration (ECS) & $[\mathrm{K^+}]_e^0$        &    3  & mM & \cite{oyehaug2012dependence}* \\
    Initial neuronal chloride concentration  (ICS) & $[\mathrm{Cl^-}]_{i, n}^0$  &    5  & mM & \cite{neuroscienceBook}* \\
    Initial glial    chloride concentration  (ICS) & $[\mathrm{Cl^-}]_{i, g}^0$  &    5  & mM & \cite{oyehaug2012dependence}* \\
    Initial          chloride concentration  (ECS) & $[\mathrm{Cl^-}]_e^0$       &  134  & mM & \cite{oyehaug2012dependence}* \\
    Initial neuronal membrane potential & $\phi_{m, n}^0$           & $-70$ & mV & \cite{sterratt2023principles}* \\
    Initial glial    membrane potential & $\phi_{m, g}^0$           & $-85$ & mV & \cite{oyehaug2012dependence}*  \\
    Resting          membrane potential & $\phi_{\mathrm{resting}}$ & $-65$ & mV & \\
    Membrane capacitance & $C_m$ & 0.01 & $\mathrm{F m^{-2}}$ & \cite{oyehaug2012dependence}\\
    Initial potassium activation probability & $n_0$ & 0.244 & & * \\
    Initial sodium activation probability    & $m_0$ & 0.029 & & * \\
    Initial sodium inactivation probability  & $h_0$ & 0.754 & & * \\
    Sodium    maximum conductivity        & $\overline{g}_{\mathrm{Na}}$ & 1200 & $\mathrm{Sm^{-2}}$ & \cite{hodgkin1952quantitative} \\
    Potassium maximum conductivity        & $\overline{g}_{\mathrm{K}}$  & 360  & $\mathrm{Sm^{-2}}$ & \cite{hodgkin1952quantitative} \\
    Neuronal sodium leak conductivity     & ${g}_{\mathrm{Na}, \mathrm{leak}, n}$   & 0.3 & $\mathrm{Sm^{-2}}$ & \cite{Benedusi2024Scalable}\textdagger \\
    Glial    sodium leak conductivity     & ${g}_{\mathrm{Na}, \mathrm{leak}, g}$   & 1 & $\mathrm{Sm^{-2}}$ & \cite{ostby2009astrocytic} \\
    Neuronal potassium leak conductivity  & ${g}_{\mathrm{K}, \mathrm{leak}, n}$    & 0.1 & $\mathrm{Sm^{-2}}$ & \cite{oyehaug2012dependence}\textdagger \\
    Glial    potassium leak conductivity  & ${g}_{\mathrm{K}, \mathrm{leak}, g}$    & 16.96 & $\mathrm{Sm^{-2}}$ & \cite{ostby2009astrocytic} \\
    Neuronal chloride leak conductivity   & ${g}_{\mathrm{Cl}, \mathrm{leak}, n}$. & 0.25 & $\mathrm{Sm^{-2}}$ & \cite{ostby2009astrocytic} \\
    Glial    chloride leak conductivity   & ${g}_{\mathrm{Cl}, \mathrm{leak}, g}$   & 2.0 & $\mathrm{Sm^{-2}}$ & \cite{ostby2009astrocytic}\textdagger \\
    Neuronal ATP pump maximum   strength  & $S_{\mathrm{ATP}, n}$   &          0.25 & $\mathrm{Am^{-2}}$ & \cite{kager2000simulated, ellingsrud2025splitting}\textdagger \\
    Glial    ATP pump maximum   strength  & $S_{\mathrm{ATP}, g}$   &          0.12 & $\mathrm{Am^{-2}}$ & \cite{Halnes2013Electrodiffusive}\textdagger \\
    ATP pump sodium    threshold          & $P_{\mathrm{Na}}$       &            10 & mM & \cite{Halnes2013Electrodiffusive} \\
    ATP pump potassium threshold          & $P_{\mathrm{K}}$        &           1.5 & mM & \cite{Halnes2013Electrodiffusive} \\
    Neuronal NKCC1 maximum strength       & $S_{\mathrm{NKCC1}, n}$ &        0.0023 & $\mathrm{Am^{-2}}$ & \cite{ellingsrud2020cell}\textdagger \\
    Glial    NKCC1 maximum strength       & $S_{\mathrm{NKCC1}, g}$ & \num{5.2e-04} & $\mathrm{Am^{-2}}$ & \cite{ostby2009astrocytic} \\
    KCC2  maximum strength                & $S_{\mathrm{KCC2}}$     &        0.0068 & $\mathrm{Am^{-2}}$ & \cite{ellingsrud2020cell}\textdagger \\
    KCC1  maximum strength                & $S_{\mathrm{KCC1}}$     &         0.018 & $\mathrm{Am^{-2}}$ & \cite{ostby2009astrocytic} \\
    Stimulus time constant                & $a_{\mathrm{stim}}$     &   \num{5e-04} & s & \\
    Maximum stimulus conductance          & $\overline{g}_{\mathrm{stim}}$ & 60--500 & $\mathrm{Sm^{-2}}$ & \\
    \bottomrule    
    \end{tabular}
\end{table}

\section{Discretization and numerical solution of the KNP-EMI equations}

The KNP-EMI model equations can be summarized as follows:
find the ionic concentrations $c_{k, r} = c_{k, r}(\xx, t)$
and electric potentials $\phi_r = \phi_r(\xx, t)$ in the
domains $\Omega_r$ such that for all $t\in[0, T]$
\begin{subequations}
\begin{alignat}{2}
    \pdifft{c_{k, r}} + \nabla\cdot\JJ_{k, r} &= f_{k, r} &&\text{in } \Omega_r,
    \label{eq:KNP_strong_concentration} \\
    F\sum_{k=1}^K z_k \nabla\cdot\JJ_{k, r} &= 0 
    &&\text{in } \Omega_r,
    \label{eq:KNP_strong_electroneutrality} \\
    \JJ_{k, r}\cdot\nn_r &= \pm_r\frac{I_{\mathrm{ion}, k} +
    \alpha_{k, r}C_m\pdifft{\phi_m}}{F z_k} 
    &&\text{on } \Gamma,
    \label{eq:KNP_strong_membrane_flux} \\
    \JJ_{k, r} &= -D_{k, r}\nabla c_{k, r} -
    \frac{z_{k, r} D_{k, r} F}{RT} c_{k, r} \nabla\phi_r \quad &&\text{in }\Omega_r,
    \label{eq:KNP_strong_nernst_planck} \\
    c_{k, r}(\xx, 0) &= c_{k, r}^0 
    &&\text{in } \Omega_r, \\
    \phi_r(\xx, 0) &= \phi_r^0 
    &&\text{in }\Omega_r, \\
    \JJ_{k, r}\cdot\nn_r &= 0 
    &&\text{on }\partial\Omega_r.
\end{alignat}
\end{subequations}
We solve these equations using the solution algorithm presented by Benedusi et al.~\cite{Benedusi2024Scalable}. The scheme is based on a finite difference method in time, an operator splitting scheme decoupling the ODEs from the PDEs, a finite element spatial discretization, and finally an iterative, preconditioned solver for the resulting linear system. For completeness, we describe the solution steps in detail below.

\subsection{Operator splitting and time discretization}

The KNP-EMI equations are nonlinear, involving both a bilinear
coupling of each concentration field with the electric potential in
each bulk domains $\Omega_r$ cf.~\eqref{eq:KNP_strong_nernst_planck}
and nonlinear dynamics at the membrane
cf.~\eqref{eq:KNP_strong_membrane_flux}. The bulk and membrane
dynamics occur on different timescales, with much faster membrane
dynamics. To decouple across these different scales, we consider an
implicit-explicit operator splitting
scheme~\cite{ellingsrud2020finite, Benedusi2024Scalable}. To this end,
we first partition the time interval $[0, T]$ uniformly into
times\footnote{Since the unknown variables already have an ion index
$k$ and a domain index $r$, we denote the timestep $n$ by a
superscript to avoid a triple subscript.}  $t^n = n\Delta t$ with $n =
1,\dots, N_t$ for $N_t$ timesteps with timestep size $\Delta t$. At
each timestep $t^n$, given the membrane potential at the previous
timestep, we first solve the gating variable ODEs
\eqref{eq:gating_variable_ODE} to obtain an approximation of the ionic
currents $I_{\mathrm{ion}, k}$. Specifically, we solve the ODEs from
$t^{n-1}$ to $t^n$ using $m$ steps of a Rush-Larsen
method~\cite{rush2007practical} with a local timestep $\Delta t_{\rm
  ODE} = \Delta t/m$ (we let $m = 25$). Next, we discretize
\eqref{eq:KNP_strong_concentration}
and~\eqref{eq:KNP_strong_membrane_flux} in time using an implicit
Euler method, and solve
\eqref{eq:KNP_strong_concentration}--\eqref{eq:KNP_strong_nernst_planck}
by evaluating the just-computed ionic currents directly. We linearize
the bilinear term by evaluating the concentration in the electrical
drift term and in the coefficients $\alpha_{k, r}$ at the previous
timestep.

\subsection{Finite element spatial discretization}

Our computational meshes are composed of tessellated intracellular
domains $\mathcal{T}_i$ and extracellular space $\mathcal{T}_e$
(\Cref{fig:model_scales}). In this section, we present the finite
element discretization for just one connected intracellular
domain. This extends to $N$ disjoint intracellular domains readily, as
will be demonstrated in~\Cref{sec:physiological_simulations}. Relative
to each mesh domain $\mathcal{T}_r$, we define the finite element
space of continuous piecewise linear polynomials $V(\mathcal{T})$
. The fully discrete weak form of the "PDE part" of the KNP-EMI
problem then reads: at each time $t^n$, find the concentration $c_{k,
  r}^{n, h} \in V(\mathcal{T}_r)$ for $k \in K$ and electric
potentials $\phi_r^{n, h} \in V(\mathcal{T}_r)$ such that
\begin{multline}
    \int_{\Omega_r} \left ( c_{k, r}^{n, h} v_{k, r} 
    - \Delta t \JJ_{k, r}^{n, h} \cdot \nabla v_{k, r} \right ) \dx
    \pm_r \frac{\alpha_{k, r}^n C_m}{z_k F} 
    \int_{\Gamma} \phi_m^{n, h} v_{k, r} \ds \\
    = \int_{\Omega_r} \left ( (c_{k, r}^{n-1, h} + \Delta t f_{k, r}^n) v_{k, r} \right ) \dx - \int_{\Gamma}{g_{k, r}^{n-1, h}v_{k, r}} \ds
    \label{eq:discrete_weak_form_concentrations} 
    \end{multline}
    for all $v_{k, r} \in V(\mathcal{T}_r), k \in K, r \in \{i, e\}$, and 
    \begin{multline}
    - \Delta t \sum_{k\in K} z_k \int_{\Omega_r} \JJ_{k, r}^{n, h} \cdot \nabla w_r \dx
    \pm_r \frac{C_m}{F}\int_{\Gamma} \phi_m^{n, h} w_r \ds
    = -\int_{\Gamma} q_r^{n-1, h} w_r \ds 
    \label{eq:discrete_weak_form_KNP}
\end{multline}
for all $w_r \in V(\mathcal{T}_r)$. Here, we have introduced the short-hand
\begin{equation}
    g_{k, r}^{n, h} = \pm_r\frac{1}{z_k F}
    \left(\Delta t I_{\mathrm{ion}, k}^{n, h}
    - \alpha_{k, r}^n C_m \phi_m^{n, h}\right)
    \text{ and }
    q_r^{n, h} = \pm_r \left(\Delta t I_{\mathrm{ion}, k}^{n, h}
    - C_m \phi_m^{n, h}\right)/F.
\end{equation}

The discrete weak form induces a linear system $\vec{A}^n \vec{x}^n
=\vec{b}^n$ to be solved at every timestep $n$.  The system matrix
$\vec{A}^n$ has the following block structure:
\begin{equation}
    \vec{A}^n = \begin{bmatrix}
        \vec{A}_{c, i} &     \vec{B}_i     &        0        &          0        \\
        \vec{B}_i^T    & \vec{A}_{\phi, i} & \vec{C}_{i, e}  &   \vec{D}_{i, e}  \\
             0         &   \vec{C}_{e, i}  &  \vec{A}_{c, e} &     \vec{B}_e     \\ 
             0         &   \vec{D}_{e, i}  &   \vec{B}_e^T   & \vec{A}_{\phi, e} \\
          \end{bmatrix}^n
    \label{eq:linear_system_matrix}
\end{equation}
where $\vec{A}_{f, r}$ is the stiffness matrix for function $f$ in the
domain $r$, $\vec{B}_r$ is the membrane potential coupling stiffness
matrix, $\vec{C}_{r_1, r_2}$ is the concentration-potential coupling
block between space $r_1$ and $r_2$, and $\vec{D}_{r_1, r_2}$ are the
potential coupling blocks.  Note that $\vec{A}_{c, r}$ are block
matrices with each of the individual $\vec{A}_{c_k, i}$ stiffness
matrices on the diagonal. The blocks
in~\eqref{eq:linear_system_matrix} coupling the intra- and
extracellular spaces arise from the shared degrees of freedom on the
cellular membranes (\Cref{fig:model_scales}).

\subsection{Solving the KNP-EMI equations in dense cortical tissue geometries}
\label{subsec:iterative_solver}

We solve the linear KNP-EMI PDE system \eqref{eq:linear_system_matrix}
in a monolithic fashion using an iterative method as previously
described~\cite{Benedusi2024Scalable}.  The iterative solver uses a
Generalized Minimal Residual Method (GMRES)~\cite{saad1986gmres} and
is preconditioned with an Algebraic Multigrid (AMG)
method~\cite{yang2002boomeramg,hypre}. We use
\begin{equation*}
    \vec{P} = \begin{bmatrix}
          \vec{A}_{c, i} &      0      &     0     &      0      \\
             0     & \vec{A}_{\phi, i} &     0     &      0      \\
             0     &      0      &  \vec{A}_{c, e} &      0      \\ 
             0     &      0      &     0     & \vec{A}_{\phi, e} \\
          \end{bmatrix}
\end{equation*}
as a preconditioner for the linear system, where all of the stiffness
matrices are evaluated with the initial conditions. We noted
previously that our choice of boundary conditions leaves the electric
potentials only determined up to a constant. Thus, without additional
constraints, the matrix $\vec{A}^n$ \eqref{eq:linear_system_matrix} is
singular. We address this by providing the iterative solver with the
nullspace of $\vec{A}^n$, which is the set of all constants. We also
orthogonalize $\vec{b}^n$ with respect to the nullspace.

The computational model is implemented in
DOLFINx~\cite{dolfinx2023preprint}, which is interfaced with the
linear algebra library PETSc. The software is parallelized with MPI
communication. The model implementation was verified in two ways
(\Cref{sec:appendix}): (i) a convergence study on idealized
geometries; (ii) a convergence study on the mouse visual cortex
meshes. The implementation is openly available at Github:
\href{https://github.com/hherlyng/knp-emi-cgx}{https://github.com/hherlyng/knp-emi-cgx}.

\begin{figure}
    \centering
    \includegraphics[width=0.8\textwidth]{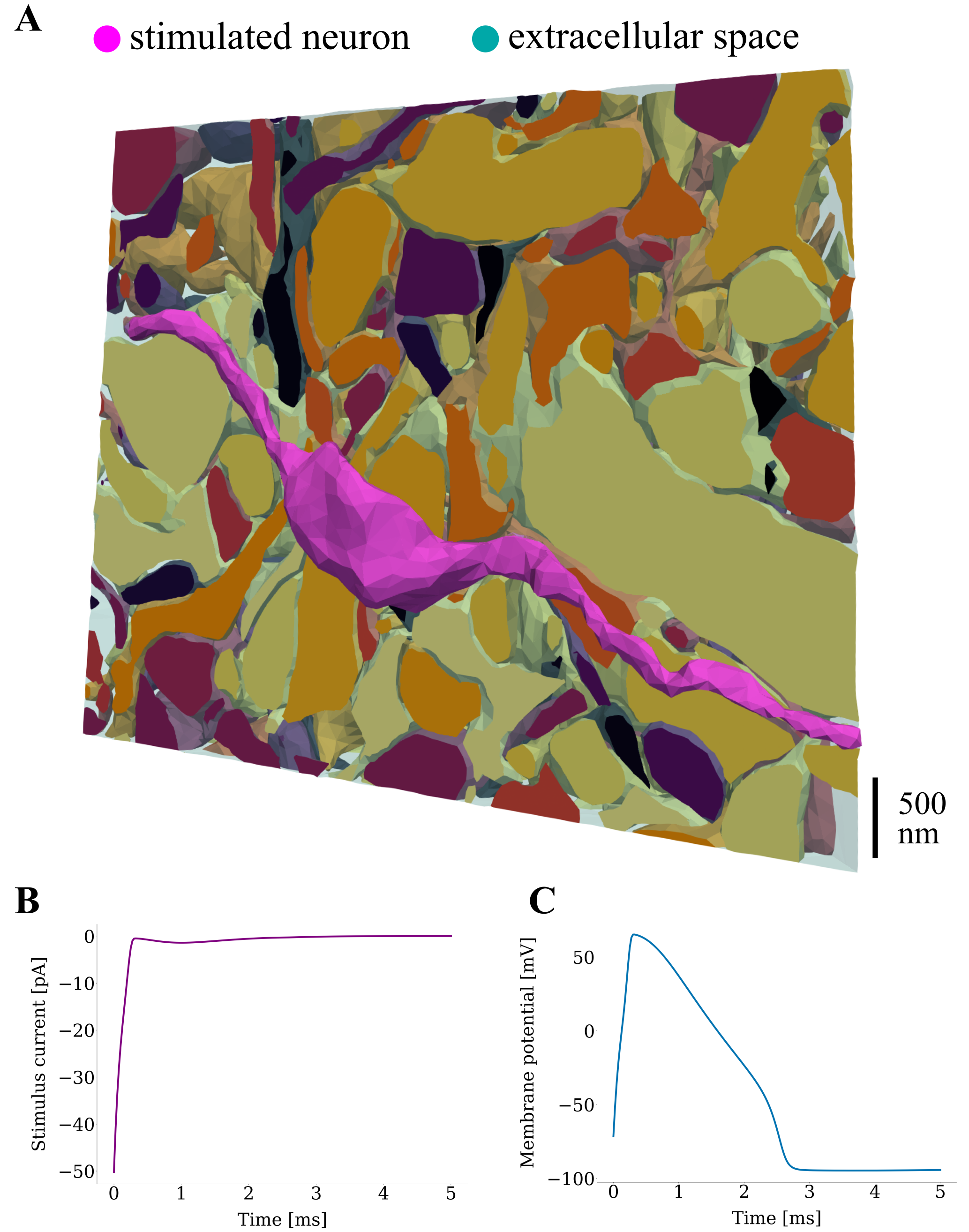}
    \caption{Simulation setup to test the performance of the iterative linear solver.
    \textbf{A.} A clip of the $L=5$ $\mu$m, $N=100$ geometry with a central
    axon highlighted in pink. The extracellular space is shown in light
    blue, while the remaining cells or cell fragments are shown in separate, unique colors.
    \textbf{B.} The total stimulus current applied to the membrane of the axon
    through adding a sodium stimulus current to the ionic membrane current density.
    \textbf{C.} The membrane potential measured in a point on the axon's membrane ($\xx = (1186, 3241, 3594)$ nm).
    }
    \label{fig:iterative_study_setup}
\end{figure}

The key question now is how this numerical solution strategy performs
in realistic or near realistic simulation scenarios defined over dense
cortical tissue geometries. To study this, we simulate the firing of
an action potential within the cortical tissue geometries generated
in~\Cref{sec:reconstruction}. In each geometry, the axon of one of the
neurons is stimulated by a stimulus current density as introduced in
\Cref{subsec:membrane_mechanisms_and_stimuli}. As an example of the
setup, \Cref{fig:iterative_study_setup}A shows the geometry for the
mesh with $L=5$ $\mu$m and $N=100$, with the stimulated axon in
pink. For simplicity in this performance test case, we consider all
other intracellular domains within the tissue cube to also be neurons,
by employing the neuronal membrane mechanisms (as introduced in
\Cref{subsec:membrane_mechanisms_and_stimuli}).

The simulations are run for a total of 200 timesteps with $\Delta t =
\num{2.5e-5}$ s to a final time of 5 ms. The total stimulus current
applied on the neuronal membrane (\Cref{fig:iterative_study_setup}B)
makes the membrane potential cross the threshold required for firing a
single action potential (\Cref{fig:iterative_study_setup}C).  We note
that the simulations are run with a varying amount of processes for
varying $L$ to achieve reasonable runtimes as the problem sizes
increase (\Cref{tab:problem_sizes}). All simulations were run using
the high-performance computing architecture eX3~\cite{ex3} on a node
with two AMD EPYC Genoa-X 9684X 96-Core dualprocessors.
\begin{table}
    \caption{Problem sizes (numbers of degrees of freedom)
    on the mouse visual cortex meshes with varying domain boundary
    side lengths $L$ and total number of biological cells $N$.
    The -- entries signify that there are
    no meshes for $L=5$ $\mu$m and $N>200$.}
    \label{tab:problem_sizes}  
    \centering
    \begin{tabular}{
        p{1cm}
        >{\raggedleft\arraybackslash}p{2.0cm}
        >{\raggedleft\arraybackslash}p{2.0cm}
        >{\raggedleft\arraybackslash}p{2.0cm}
        >{\raggedleft\arraybackslash}p{2.0cm}
        >{\raggedleft\arraybackslash}p{2.0cm}
        >{\raggedleft\arraybackslash}p{2.0cm}
    }
    \toprule
    $L$ [$\mu$m] & $N=100$ & $N=200$ & $N=300$ & $N=400$ & $N=500$ \\
    \midrule
    5 &  1 219 172 &  1 452 136 &     --     &     --     &     --     \\
    10 &  4 161 248 &  5 683 448 &   8 926 676 &   7 401 376 &  10 729 424 \\
    20 & 22 892 896 & 34 403 796 &  31 837 296 &  35 739 444 &  48 709 344 \\
    30 & 57 002 756 & 78 171 324 & 105 776 564 & 121 689 728 & 111 075 140 \\
    \bottomrule
    \end{tabular}
\end{table}    

We first examine the total assembly times -- the total time spent
assembling the system matrix, right-hand side vector and
preconditioner matrix over the whole simulation time
(\Cref{tab:assembly_times}). As expected, these times increase with
increasing problem size in a controlled fashion. A general trend is
that assembly times remain fairly constant when concomitantly doubling
the number of MPI cores and the number of degrees of freedom,
indicating scalable assembly.
\begin{table}
    \caption{Assembly times when simulating an action potential on the
      mouse visual cortex meshes with varying domain boundary side
      lengths $L$ and total number of biological cells $N$.  The --
      entries signify that there are no meshes for $L=5$ $\mu$m and
      $N>200$.}
    \label{tab:assembly_times}  
    \centering
    \begin{tabular}{
        p{2.1cm}
        >{\raggedleft\arraybackslash}p{1.45cm}
        >{\raggedleft\arraybackslash}p{1.45cm}
        >{\raggedleft\arraybackslash}p{1.45cm}
        >{\raggedleft\arraybackslash}p{1.45cm}
        >{\raggedleft\arraybackslash}p{1.45cm}
        >{\raggedleft\arraybackslash}p{1.45cm}
    }
    \toprule
    $L$ [$\mu$m] & $N=100$ & $N=200$ & $N=300$ & $N=400$ & $N=500$ \\
    \midrule
    5  (16 cores) & 1390 s &  1783 s &   -- &      -- &     -- \\ 
    10  (32 cores) & 2269 s &  3459 s &  5543 s &     5414 s &    8314 s \\
    20  (64 cores) & 7219 s & 10749 s & 10260 s &    12751 s &   16355 s \\
    30 (128 cores) & 7597 s & 14828 s & 21652 s &  &   \\
    \bottomrule
    \end{tabular}
\end{table}

\begin{figure}
    \centering
    \includegraphics[width=\textwidth]{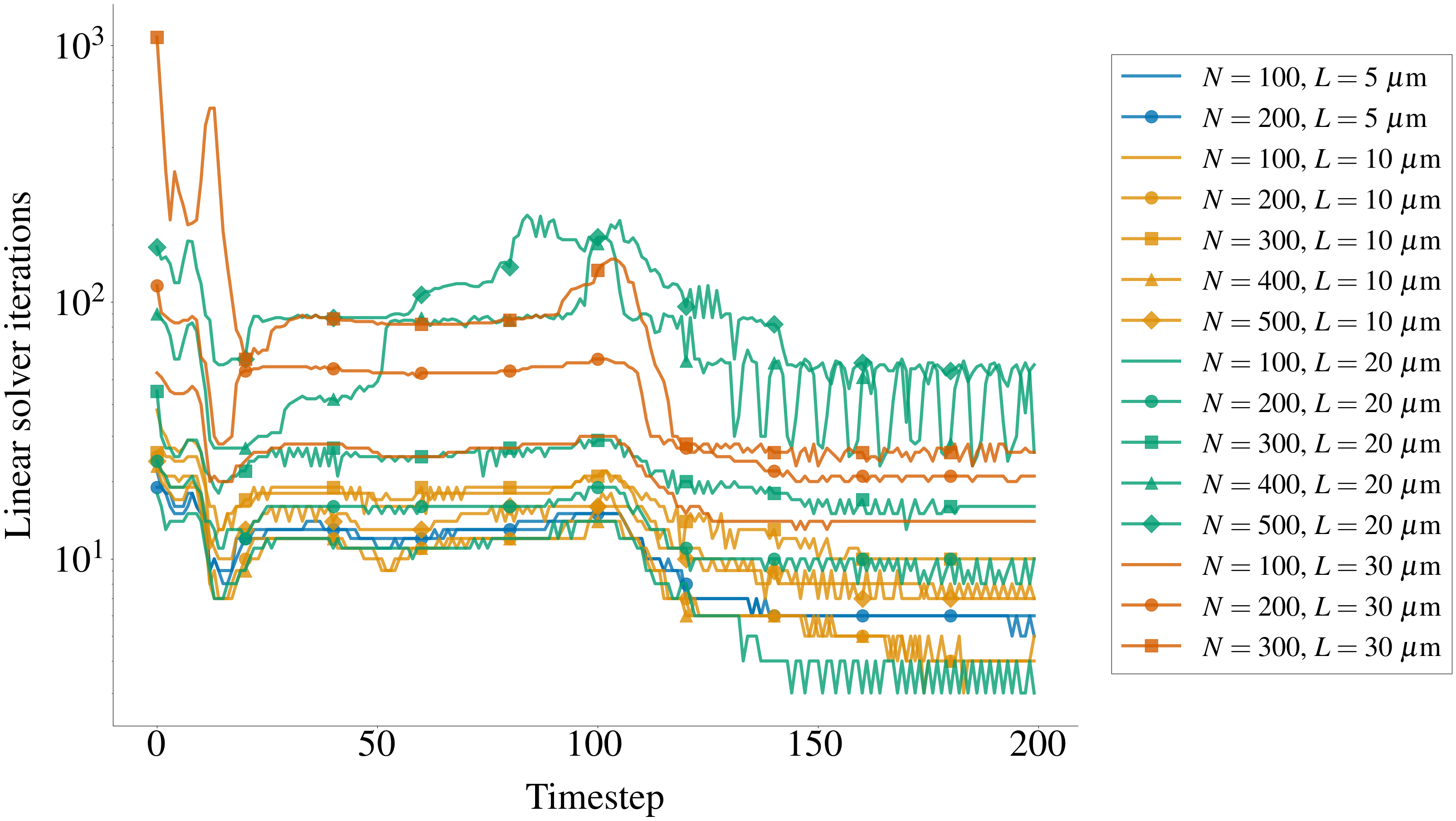}
    \caption{Iteration counts for 200 timesteps when solving the
      KNP-EMI equations on the various mouse visual cortex geometries
      during a simulation of an action potential.  Note the
      logarithmic scaling of the vertical axis.  }
    \label{fig:iterations}
\end{figure}

Turning to look at per-timestep linear solver iterations until
convergence, the number of iterations is highest during the action
potential when the variables rapidly change with time
(\Cref{fig:iterations}).  After the membrane potential peak around
timestep 25, iteration counts remain fairly constant until a slight
increase concurrent with the membrane potential kink at around
timestep 100. For the last 100 iterations the iteration counts fall at
first before stabilizing, as a result of the membrane potential being
fairly constant.

The solver performance in terms of average iterations is stable for
the smaller geometries with $L=5$ $\mu$m and $L=10$ $\mu$m, and for
the $L=20$ $\mu$m geometries with up to $N=300$ biological cells,
while the iteration counts start increasing from $N=300$ towards
$N=500$ (\Cref{tab:solver_times_and_iterations}). For the $L=30$
$\mu$m geometries, iteration counts are higher than the smaller
geometries and increase with increasing $N$. For $N=400$ and $N=500$,
the linear solver reached the prescribed maximum number of iterations
per timestep (5000). The total time spent solving the linear system
generally increases as problem sizes increase.
\begin{table}
    \caption{Solver times (in seconds) and average iteration counts
      (in brackets) when simulating an action potential on the mouse
      visual cortex meshes with varying domain boundary side lengths
      $L$ and total number of biological cells $N$. The -- entries
      signify that there are no meshes for $L=5$ $\mu$m and $N>200$.
      Max its. means the simulation exceeded the prescribed maximum
      number of linear solver iterations for a single timestep
      (5000).}
    \label{tab:solver_times_and_iterations}  
    \centering
    \begin{tabular}{
        p{2.35cm}
        >{\raggedleft\arraybackslash}p{1.65cm}
        >{\raggedleft\arraybackslash}p{1.65cm}
        >{\raggedleft\arraybackslash}p{1.65cm}
        >{\raggedleft\arraybackslash}p{1.65cm}
        >{\raggedleft\arraybackslash}p{1.65cm}
        >{\raggedleft\arraybackslash}p{1.65cm}
    }
    \hline\noalign{\smallskip}
    $L$ [$\mu$m] & $N=100$ & $N=200$ & $N=300$ & $N=400$ & $N=500$ \\
    \noalign{\smallskip}\svhline\noalign{\smallskip}
     5  (16 cores) &  175 s [10] &   203 s [10] &    --   &     --    &    --      \\
    10  (32 cores) &  435 s [14] &   443 s [9] & 1103 s [16] &  648 s [9] &  1263 s [12] \\
    20  (64 cores) & 1105 s [9] &  2316 s [13] & 3258 s [22] & 10400 s [62] & 21172 s [100] \\
    30 (128 cores) & 3206 s [22] & 10901 s [40] & 30127 s [86] &  max its. &  max its. \\
    \noalign{\smallskip}\hline\noalign{\smallskip}
    \end{tabular}
\end{table}

We now look at strong scaling of the solver by comparing performance
on the $L=10$ $\mu$m geometries when using 16 and 32 cores. Assembly
times are roughly cut in half when doubling the number of
cores~(\Cref{tab:scaling_study}). The solver times are halved for the
$N=100$ and $N=400$ geometries, and even more than halved for the
$N=200$ geometry. Mostly, the average number of iterations are similar
when running on 16 and 32 cores. An interesting exception is the
$N=300$ geometry, for which the average iteration count almost doubles
when doubling the number of cores.
\begin{table}
    \caption{Solver times and average iteration counts when simulating
      an action potential on the mouse visual cortex meshes with the
      $L=10$ $\mu$m geometries for varying number of biological cells
      $N$.}
    \label{tab:scaling_study}  
    \centering
    \begin{tabular}{
        p{3cm}
        >{\raggedleft\arraybackslash}p{1.55cm}
        >{\raggedleft\arraybackslash}p{1.55cm}
        >{\raggedleft\arraybackslash}p{1.55cm}
        >{\raggedleft\arraybackslash}p{1.55cm}
        >{\raggedleft\arraybackslash}p{1.55cm}
        >{\raggedleft\arraybackslash}p{1.55cm}
    }
    \toprule
    Metric (\# Cores) & $N=100$ & $N=200$ & $N=300$ & $N=400$ & $N=500$ \\
    \midrule
        Assembly (16) & 4796 s & 7621 s & 12354 s & 10745 s & 15819 s \\
    Assembly (32) & 2269 s & 3459 s &  5543 s &  5414 s &  8314 s \\ 
    Solve (16) &  844 s & 1283 s & 1443 s & 1204 s & 2578 s \\
    Solve (32) &  435 s &  443 s & 1103 s &  648 s & 1263 s \\
    Average iterations (16) & 13 & 13 &  9 & 9 & 15 \\
    Average iterations (32) & 14 &  9 & 16 & 9 & 12 \\
    \bottomrule
    \end{tabular}
\end{table}

In summary, these findings indicate that the solution strategy
performs well for reasonably complex meshes, but also that there are
clear opportunities for further research in solution algorithms for
simulation of electrodiffusion in dense tissue reconstructions.

\section{Electrodiffusive interplay in geometrically detailed brain tissue during low- and high-frequency firing}
\label{sec:physiological_simulations}

Neuronal activity is governed by ionic movement across cellular
membranes. A key change is the release of potassium from excited
neurons into the extracellular space (ECS), which, if not cleared
efficiently, can enhance neuronal excitability and disrupt neuronal
function~\cite{aitken1986sources, nicholson1978calcium,
  utzschneider1992mutual}. The restoration of extracellular ionic
levels, via e.g.~glial uptake or homeostatic mechanisms such as the
$\text{Na}^+/\text{K}^+$-$\text{ATPase}$ sodium-potassium pump,
following neuronal activity is thus critical for healthy brain
function~\cite{kofuji2004potassium, walz2000role}. During sustained
high-frequency firing, the homeostatic mechanism are not able to keep
up, leading to larger, persistent ionic shifts that will alter
excitably and firing patterns~\cite{somjen2002ion}. The intricate
morphology of individual cells and the surrounding ECS may further
affect local electrodiffusive dynamics, a factor often neglected in
simplified models.

\begin{figure}
    \centering
    \includegraphics[width=\textwidth]{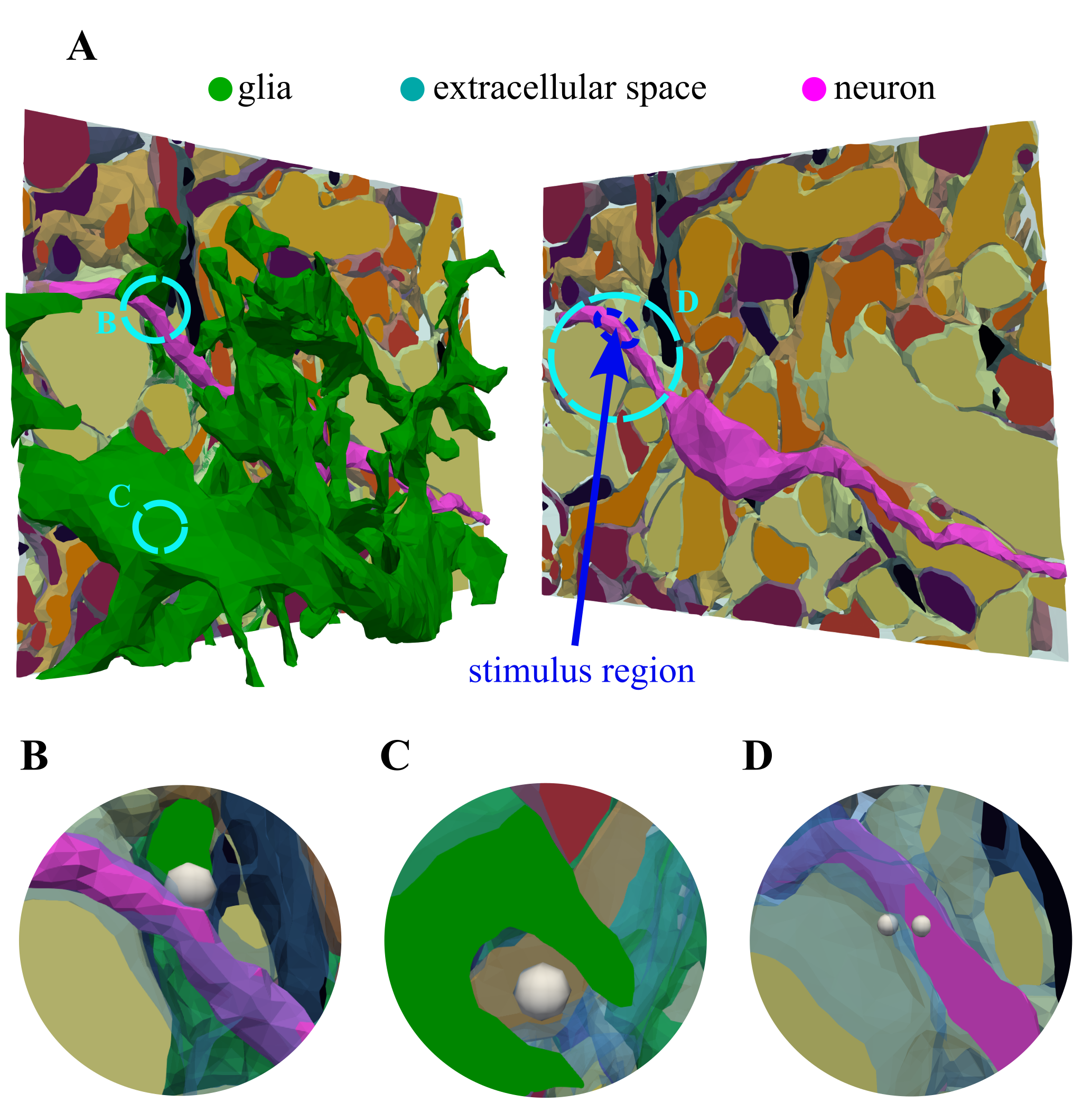}
    \caption{
    \textbf{A.} Illustration of the computational mesh with $L=5$ $\mu$m
    and $N=100$ with an astrocyte (green) partly
    wrapping around the stimulated axon (pink). The dashed blue ellipse indicates the stimulus region.
    The extracellular space is colored opaque light blue.
    All other colored structures are biological cells, with a unique
    color for every unique cell.
    \textbf{B.} Gray spherical glyph denoting point where glial intracellular
    concentrations are measured.
    \textbf{C.} Gray spherical glyph denoting point where distal extracellular
    concentrations are measured.
    \textbf{D.} Gray spherical glyphs denoting points where proximal extracellular
    concentrations (left) and neuronal intracellular concentrations (right)
    are measured.
    }
    \label{fig:physiological_study_setup}
\end{figure}

To demonstrate the capabilities of the electrodiffusive simulation
framework detailed in the present work, we consider the two scenarios
(i) low-frequency (normal) neuronal activity and (ii) high-frequency
firing, in a geometrically detailed domain with neurons, glial cells
and ECS. We analyze and compare the two scenarios and point to
observations where shape affect the dynamics (and thus potentially
function). Specifically, we consider the mesh introduced
in~\Cref{sec:reconstruction} with a domain length $L = 5$ $\mu$m and
$N = 100$ biological cells.  The membrane model is described
in~\Cref{subsec:membrane_mechanisms_and_stimuli} and notably includes
homeostatic mechanisms. We further apply the stimulus
current~\eqref{eq:stimulus_current_density} and consider two different
stimuli with periods $T_{\mathrm{stim}} = 1$ s and $T_{\mathrm{stim}}
= 20$, which correspond to firing frequencies of 1 and 50 Hz,
respectively.  The stimulus current is applied on all mesh facets of
the axon where $x\in (1.15, 1.25)$ $\mu$m. The tissue domain is
illustrated in~\Cref{fig:physiological_study_setup}.

\subsection{Membrane potentials and ion concentrations develop over different time scales during low neuronal activity
        }
\label{subsec:simulating_low_neuronal_activity}

\begin{figure}
    \centering
    \includegraphics[width=0.925\textwidth]{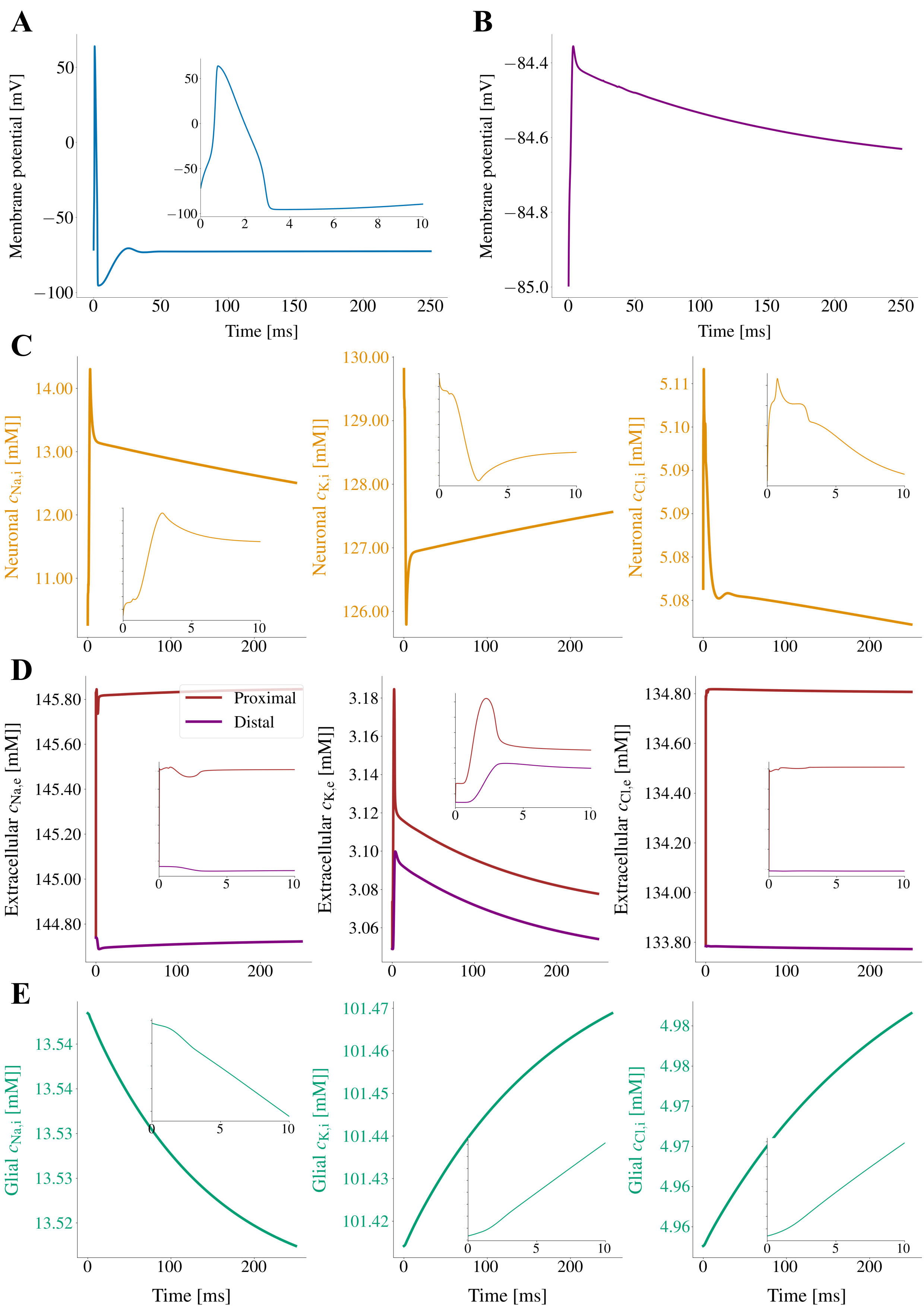}
    \caption{Neuronal membrane potential (\textbf{A}), glial membrane potential (\textbf{B}),
    neuronal ion concentrations (\textbf{C}), extracellular ion concentrations (\textbf{D})
    and glial ion concentrations (\textbf{E})
    when simulating stimulation of an axon at a physiological frequency
    of 1 Hz on the $L=5$ $\mu$m mouse visual cortex mesh
    with $N=100$ biological cells. The insets show the graphs during the first 10
    ms. The insets have separate axes limits for the ICS and ECS concentrations (like in
    the original plots). In panel \textbf{C}, proximal and distal refer to 
    the ECS points shown in \Cref{fig:physiological_study_setup}D and C, respectively.
    }
    \label{fig:point_values_low_activity}
\end{figure}

We start by analyzing the temporal dynamics of the system during a low
firing frequency of 1 Hz over a period of $250$ ms. We observe that an
action potential fires at the onset of the simulation as a result of
the stimulus current applied to the axon
(\Cref{fig:point_values_low_activity}A). After the action potential,
the membrane potential returns to a resting state at $-72.5$ mV, which
is a 1.4\% change from the initial state of $-71.5$ mV
(\Cref{fig:point_values_low_activity}A). Meanwhile, the glial membrane
potential initially increases rapidly during the action potential, by
up to~$0.6$ mV from the initial value of $-85$ mV, before it steadily
decreases over time reaching~$-84.6$ mV at $250$ ms
(\Cref{fig:point_values_low_activity}B).

The action potential initiates rapid, coupled ionic shifts in both the
intracellular and extracellular compartments. The neuronal $\rm{Na}^+$
and $\rm{K}^+$ concentrations respectively increase and decrease by
approximately 4 mM during the action potential, before the
concentrations slowly go back toward baseline levels as the
homeostatic mechanisms (pumps, co-transporters and glial uptake) are
activated. We observe that the neuronal $\rm{Cl}^-$ concentration
rapidly increases as the neuron depolarizes, and decrease as the
neuron repolarizes. Moreover, the $\rm{Cl}^-$ concentration drops
below the initial state, indicating a total loss of neuronal
$\rm{Cl}^-$. Note that the changes in $\rm{Cl}^-$ are two orders of
magnitude smaller than those observed for $\rm{Na}^+$ and $\rm{K}^+$
(\Cref{fig:point_values_low_activity}C).

In the extracellular space proximal to the stimulated region, all ion
concentrations change abruptly when the stimulus current is applied
(\Cref{fig:point_values_low_activity}D).  In contrast to the neuronal
concentrations, the shifts in the extracellular concentrations are of
similar magnitude for all species.  After the action potential is
fired, both extracellular $\text{Na}^{+}$ and $\text{Cl}^{-}$
concentrations proximal to the stimulated neuron settle at elevated
values (around $1$ mM higher than the initial concentrations). The
extracellular $\text{K}^{+}$ concentration increases during the action
potential, before decreasing as the homeostatic mechanism are
activated, and is still decreasing at the end of the
simulation. Distal to the neuron, the extracellular $\text{Na}^{+}$
and $\text{Cl}^{-}$ concentrations are decreased by $0.017$ and
$0.012$ mM, respectively, an order of magnitude less than the shifts
of these ions proximal to the neuron. The distal $\text{K}^{+}$
concentration behaves similar to the proximal concentration, but the
increase from the initial $\text{K}^{+}$ concentration to the peak
concentration value in the distal point is half of that observed in
the proximal point (\Cref{fig:point_values_low_activity}D).

The glial ion concentrations develop steadily over time throughout the
entire simulation, with a decreasing $\text{Na}^{+}$ concentration
while $\text{K}^{+}$ and $\text{Cl}^{-}$ concentrations increase
(\Cref{fig:point_values_low_activity}E).  The ion concentration
changes of the three species are similar in magnitude; at the end of
the simulation, the $\text{Na}^{+}$ concentration has decreased by
0.026 mM, while $\text{K}^{+}$ and $\text{Cl}^{-}$ concentrations have
increased by 0.055 and 0.029 mM, respectively. In general, the ion
concentrations stabilize over a much slower time scale than the
membrane potentials. Both the membrane potentials and the
concentrations approach a different set of steady-state values than
the initial conditions (\Cref{fig:point_values_low_activity}C--E).

\subsection{Sustained high-frequency firing induces a burst-firing mode}

\begin{figure}
    \centering
    \includegraphics[width=0.925\textwidth]{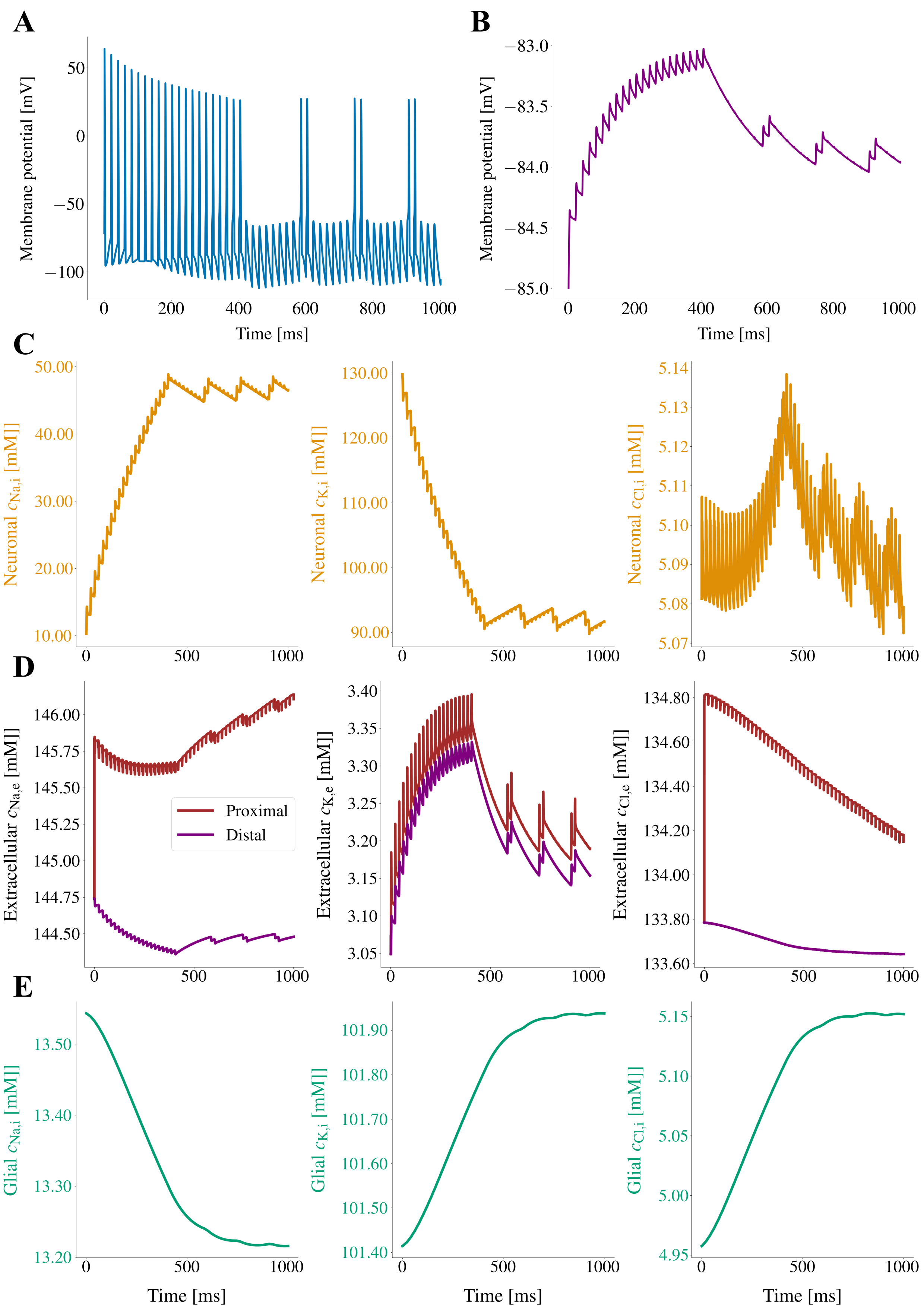}
    \caption{Neuronal membrane potential (\textbf{A}), glial membrane
      potential (\textbf{B}), neuronal ion concentrations
      (\textbf{C}), extracellular ion concentrations (\textbf{D}) and
      glial ion concentrations (\textbf{E}) when simulating
      stimulation of an axon at a frequency of 50 Hz on the $L=5$
      $\mu$m mouse visual cortex mesh with $N=100$ biological
      cells. The insets show the graphs during the first 10 ms. The
      insets have separate axes limits for the ICS and ECS
      concentrations (like in the original plots). In panel
      \textbf{C}, proximal and distal refer to the ECS points shown in
      \Cref{fig:physiological_study_setup}\textbf{D} and \textbf{C},
      respectively.}
    \label{fig:point_values_hyperactivity}
\end{figure}

During high-frequency firing, where the axon is stimulated at a
frequency of $50$ Hz, the neuron initially fires a train of action
potentials for approximately $400$ ms
(\Cref{fig:point_values_hyperactivity}A). At $t = 400$ ms, the neuron
transitions into a new firing mode, characterized by bursts of two
action potentials separated by silent periods that last around 180 ms.
We observe that the amplitude of the membrane depolarization decreases
over the first 400 ms; the initial AP peaks at 65 mV, while the last
AP before the neuron transitions into a new firing mode peaks at less
than 28 mV. We further observe an alteration of the hyperpolarization
phase for each AP during the initial firing phase. Correspondingly,
the glial membrane potential exhibits a slow, gradual depolarization
over the $1000$ ms simulation time, overlaid with small, rapid
oscillations corresponding to the neuronal APs
(\Cref{fig:point_values_hyperactivity}B). Notably, the initial AP
depolarizes the glial membrane more heavily than the following APs.

The transition in neuronal firing pattern is underpinned by
significant ionic shifts in both the intracellular and extracellular
compartments. We observe that the neuronal $\text{Na}^{+}$
concentration increases rapidly while the $\text{K}^{+}$ concentration
decreases rapidly during the initial firing phase. Both
concentrations change by almost 40 mM during the sustained train of
APs in the first 400 ms of the simulation. Following the transition
to the burst-firing mode, the ion concentration trends reverse during
the silenced periods, indicating that the homeostatic mechanisms are
activated. The neuronal $\text{Cl}^{-}$ concentration displays
significant oscillations, peaking near the onset of the second firing
phase (\Cref{fig:point_values_hyperactivity}C). The $\text{Cl}^{-}$
concentration changes are two orders of magnitude smaller then
$\text{Na}^{+}$ and $\text{K}^{+}$ concentration changes.

We further observe that the extracellular $\text{K}^{+}$ concentration
increases rapidly during the initial firing phase, reaching a peak of
approximately $3.40$ mM and $3.33$ mM in the proximal and distal
points, respectively, before decreasing (with notable peaks during the
bursts) in the second firing phase
(\Cref{fig:point_values_hyperactivity}D).  The extracellular
$\text{Na}^{+}$ and $\text{Cl}^{-}$ concentrations decrease
consistently over the simulation period in the distal point, whereas
in the proximal point both concentrations increase rapidly at the
onset of the simulation before $\text{Na}^{+}$ slowly increases while
$\text{Cl}^{-}$ decreases.

In contrast to the spiking behavior observed in the neuron and the
ECS, the glial ion concentrations change gradually and more smoothly
over time during both the AP train and the burst-firing mode
(\Cref{fig:point_values_hyperactivity}E).  The $\text{Na}^{+}$
concentration decreases, while both $\text{K}^{+}$ and $\text{Cl}^{-}$
increase.  The rates of change decrease after the onset of the
burst-firing mode, and all ion concentrations appear to saturate.

\subsection{Cellular morphology determines how ion concentrations and electric potentials develop in space during neuronal activity}
\begin{figure}
    \centering
    \includegraphics[width=\textwidth]{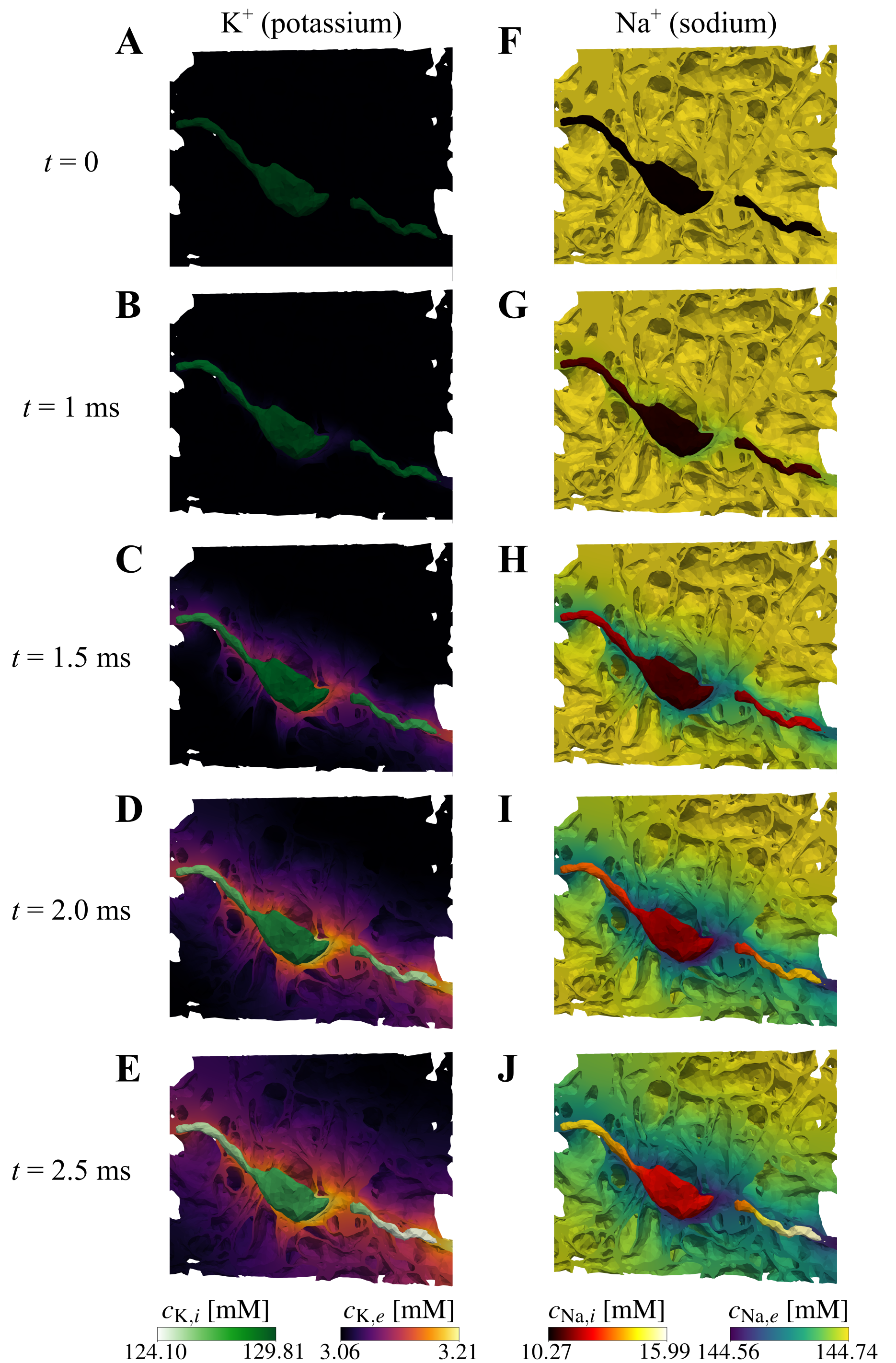}
    \caption{Potassium (\textbf{A}--\textbf{E}) and sodium 
    (\textbf{F}--\textbf{J}) concentrations in the stimulated
    neuron and the extracellular space 
    over time when simulating the firing of an
    action potential using the $L=5$ $\mu$m, $N=100$ mouse visual cortex
    geometry. The ECS is clipped and no other cells than the 
    stimulated neuron are shown.}
    \label{fig:3D_concentrations_neuron_during_action_potential}
\end{figure}

We now turn to the spatial distribution of ion concentrations and
electric potentials. During firing of a single action potential, the
extracellular $\text{K}^{+}$ concentration is fairly constant in the
initial phase when the membrane potential rises
(\Cref{fig:3D_concentrations_neuron_during_action_potential}A, B).
Subsequent to the membrane potential peak, the increase in
extracellular $\text{K}^{+}$ spreads through the extracellular space
fast as the voltage-gate $\text{K}^{+}$ channels open
(\Cref{fig:3D_concentrations_neuron_during_action_potential}C--E).
During the action potential, the extracellular $\text{Na}^{+}$
concentration decreases gradually proximal to the neuron, with less
notable changes in distal locations
(\Cref{fig:3D_concentrations_neuron_during_action_potential}F--J).

There are notable local differences in the neuronal concentrations,
with distinctly greater $\text{K}^{+}$ and lower $\text{Na}^{+}$
concentrations in the axonal varicosity (widened section) of the
stimulated neuron after the action potential, compared to the slender
structures of the axon.
(\Cref{fig:3D_concentrations_neuron_during_action_potential}E, J).  At
time $t=2.5$ ms, the $\text{Na}^{+}$ concentration is 2-3 mM greater
in the slender structures than in the varicosity.

\begin{figure}
    \centering
    \includegraphics[width=\textwidth]{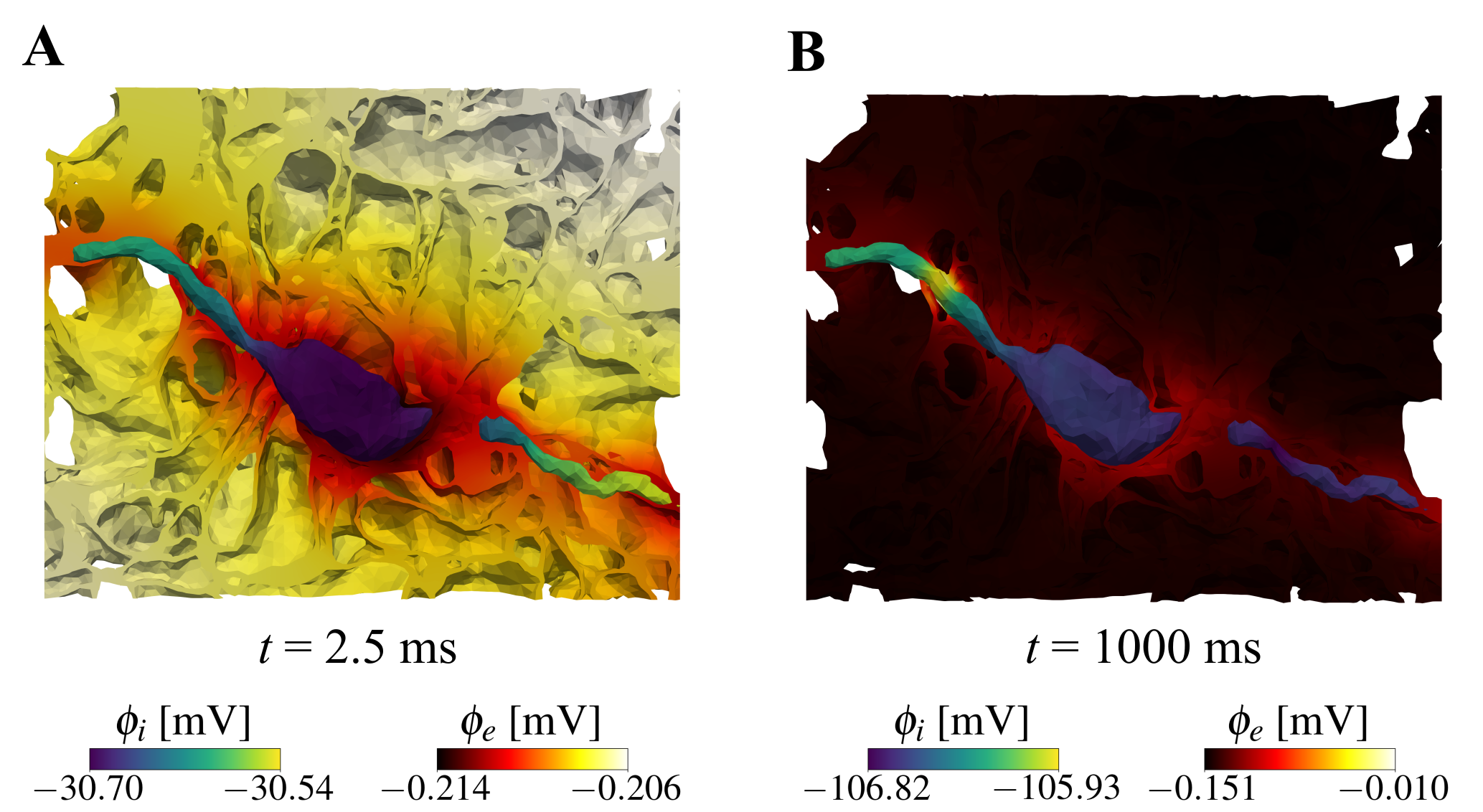}
    \caption{Electric potentials in the stimulated neuron and
    the extracellular space at times $t=2.5$ ms (\textbf{A})
    and $t=1000$ ms (\textbf{B}).
    The ECS is clipped and the only cell shown
    is the neuron that is stimulated.
    Note the individual colorbars for each of the subpanels.}
    \label{fig:3D_potentials_neuron}
\end{figure}

At a given time instant, there are also small spatial differences in
the electric potentials in the stimulated neuron and the extracellular
space.  After the peak of the first action potential at time $t = 2.5$
ms, the extracellular potential $\phi_e$ is most negative close to the
neuron, especially near the axonal varicosity
(\Cref{fig:3D_potentials_neuron}A).  The intracellular potential
$\phi_i$ is also lowest on the membrane of the varicosity. After
$t=1000$ ms, subsequent to sustained stimulus with a frequency of 50
Hz, the situation is reversed for the extracellular potential
(\Cref{fig:3D_potentials_neuron}B). Now, the potential is less
negative close to the neuron. Also, the maximum spatial differences in
$\phi_e$ at time $t=1000$ ms are around 0.14 mV in magnitude, more
than an order greater than the maximum difference of 0.008 mV at
$t=2.5$ ms.

\begin{figure}
    \centering
    \includegraphics[width=\textwidth]{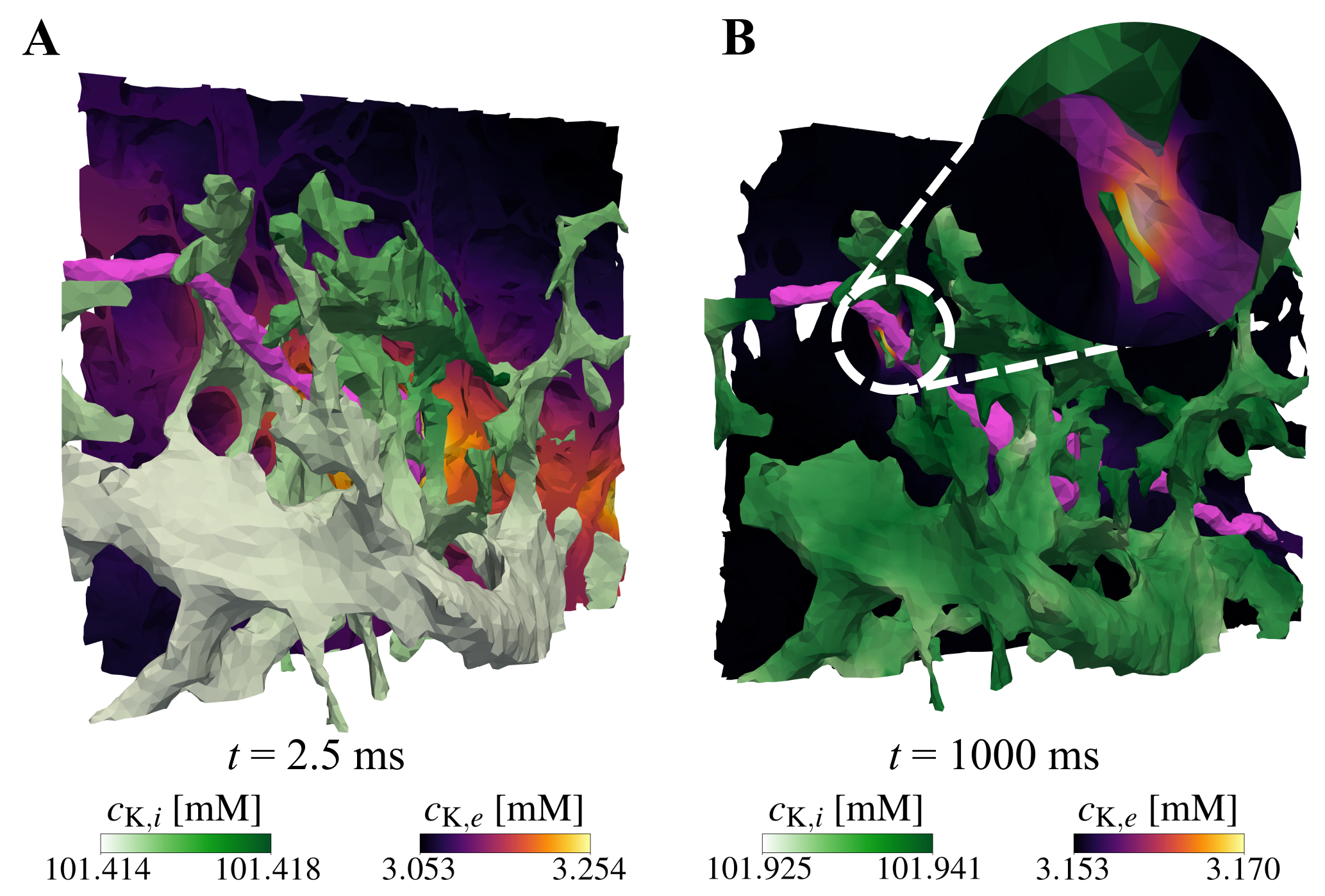}
    \caption{Potassium concentrations in an astrocyte (glial cell) and the 
    extracellular space at times $t=2.5$ ms (\textbf{A}) and $t=1000$ ms (\textbf{B})
    when simulating a neuron (pink) with a frequency of 50 Hz.
    The ECS is clipped.
    Note the individual colorbars for each of the subpanels.
    }
    \label{fig:3D_potassium_glia}
\end{figure}

In the astrocyte wrapping around the stimulated neuron, the
intracellular $\text{K}^{+}$ at $t=2.5$ ms has increased most close to
the neuron and in slender structures (\Cref{fig:3D_potassium_glia}A).
The spatial variation in glial $\text{K}^{+}$ concentration is as low
as 0.002 mM, whereas in the extracellular space, the concentrations
differ by around 0.14 mM.  Later at $t=1000$ ms, the intracellular
$\text{K}^{+}$ differs by around 0.016 mM
(\Cref{fig:3D_potassium_glia}B).  The extracellular $\text{K}^{+}$
concentration is essentially uniform in space, except for locally
elevated values in the stimulus region (\Cref{fig:3D_potassium_glia}B
inset).

\section{Discussion and outlook}

We have presented a computational framework for simulation of
electrodiffusion in cellular geometries based on a finite element
method for the KNP-EMI equations.  We have also demonstrated a
pipeline for generating realistic computational meshes from electron
microscopy data that represents dense reconstructions of cerebral
tissue.  Using such a geometry of mouse visual cortex tissue, we
applied our computational framework to simulate electrodiffusive
dynamics during neuronal activity at low- and high-frequency
firing. The results clearly show that morphology and shape affect the
electrodiffusive dynamics.

We also reported on the characteristics of dense reconstructions of
cerebral tissue by generating a series of meshes with varying domain
sizes and number of biological cells.  We found that the number of
membrane points in the geometries are high compared to the total
number of points, especially for geometries with a high number of
biological cells.  This poses a challenge for strongly coupled
problems such as the the electrodiffusive KNP-EMI equations, with ion
concentrations and electric potentials developing over varying
temporal and spatial scales.  Since theory based on spectral analysis
and simulations based on the EMI equations suggest that the
preconditioner we employ will struggle for a high ratio of membrane
points \cite{benedusi2024dense, benedusi2024modeling,
  Benedusi2024Scalable}, we assessed how the iterative solver performs
in this setting.  For the smaller geometries, the solver performed
consistently in spite of high membrane-point ratios, but performance
deteriorated with increasingly complex geometries. It would be
interesting to investigate how PDE splitting strategies, contrasting
our monolithic approach, would perform on these geometries.

In terms of limitations, two important considerations are the
following.  First, there are uncertainties in the membrane mechanism
parameters that are adopted from simpler models.  This calls for
validation by comparing simulations with physical experiments. Another
alternative is an uncertainty quantification study. Second, the
geometries used have narrow extracellular structures where the
distance separating two cells may approach the limit of the
electroneutrality assumption in the KNP equations. The KNP
approximation has shown to be excellent on the micrometer
scale~\cite{halnes2016effect}.  When modeling dynamics very close to
the cellular membranes at the nanometer scale, the small spatial and
temporal scales require solving a Poisson equation for the electric
potentials, forming the Poisson-Nernst-Planck (PNP) equations to fully
incorporate the rapid dynamics~\cite{jaeger2023nano}.  A comparison
study of the two approaches (KNP and PNP) on the mouse visual cortex
geometries could shed light on this matter.

Altogether, the content of this chapter exemplifies that scientific
computing has developed to the point where numerical solutions of
partial differential equations on realistic brain tissue geometries
are feasible.  There are however still unresolved questions, but
future works can hopefully continue to push the boundaries of
computational neuroscience studies even further.

\begin{acknowledgement}
We are grateful to Pietro Benedusi for interesting discussions and for
providing input to \Cref{fig:model_scales}. The research presented in
this chapter has benefited from the Experimental Infrastructure for
Exploration of Exascale Computing (eX3), a computational
infrastructure that is financially supported by the Research Council
of Norway under contract \#270053. G.~T.~E. acknowledges funding from
the European Union Horizon 2020 Research and Innovation Programme
under Grant Agreement No. 101147319 (EBRAINS 2.0). M.~C. acknowledges
funding from the European Research Council under grant No. 101141807
(aCleanBrain). M.E.R.~acknowledges support from Stiftelsen Kristian
Gerhard Jebsen via the K.~G.~Jebsen Centre for Brain Fluid Research
and Wellcome via Award 313298/Z/24/Z (Next-generation simulation and
learning in imaging-based biomedicine).
\end{acknowledgement}

\ethics{Competing Interests}{The authors have no conflicts of interest
  to declare that are relevant to the content of this chapter.}

\bibliographystyle{spmpsci}
\bibliography{ref}

\newpage
\section*{Appendix}
\label{sec:appendix}
\addcontentsline{toc}{section}{Appendix}
\subsection*{Numerical verification of the iterative numerical method with idealized geometries}
Here, we verify spatial convergence properties of the
iterative numerical method by performing a
convergence study with two idealized problems.
The first problem considers the
domain $\Omega = [0, 1]^2$ $\mu\mathrm{m^2}$ with boundary $\partial\Omega$,
with an inner square  $\Omega_i = [0.25, 0.75]^2$ $\mu\mathrm{m^2}$
as intracellular space and $\Omega_e = \Omega\setminus\Omega_i$
as extracellular space. The interface that separates $\Omega_i$ 
and $\Omega_e$ is the cellular membrane $\Gamma$.
The second problem is similar, and considers the cube 
$\Omega = [0, 1]^3$ $\mu\mathrm{m^3}$ with an inner cube
$\Omega_i = [0.25, 0.75]^3$ $\mu\mathrm{m^3}$
as intracellular space and $\Omega_e = \Omega\setminus\Omega_i$
as extracellular space. The cellular membrane $\Gamma$ is 
also here defined as the interface
between $\Omega_i$ and $\Omega_e$. The two problem domains
are meshed with triangles and tetrahedra with
mesh edge lengths $h$. To assess convergence,
we consider a series of increasingly refined meshes.
The coarsest mesh edge length of the two problems
is denoted $h_0$.

Convergence is verified by manufacturing solutions
for both problems~\cite{Roache2001CodeSolutions}.
The parameter values of $R$, $F$, $T$ and $C_m$ are all set to 1.
We solve the problems for one timestep using $\Delta t = \num{1e-5}$,
and compute the $L^2$-error for the concentrations and potentials
\Cref{tab:convergence_rates_2D,tab:convergence_rates_3D}.
Although a priori error estimates for the KNP-EMI problem are not yet
well-established, the convergence rate when using continuous Lagrange elements
of order $p$ to approximate a finite element solution
is bounded from above by $p+1$~\cite{johnson2009numerical}.
We observe convergence rates of $p+1$,
both for the 2D and the 3D test problems.
\begin{table}
    \centering
    \caption{Error norms and convergence rates
    for the two-dimensional verification problem
    on meshes of increasing resolution characterized
    by the edge lengths $h$.
    The coarsest mesh has mesh edge length $h_0$.
    }\label{tab:convergence_rates_2D}
    \begin{tabular}{p{0.9cm}
        >{\centering\arraybackslash}p{2.65cm}
        >{\centering\arraybackslash}p{2.65cm}
        >{\centering\arraybackslash}p{2.65cm}
        >{\centering\arraybackslash}p{2.65cm}
    }
    \hline\noalign{\smallskip}
    $h/h_0$ &
    $\|c_{\mathrm{Na}, i} - c_{\mathrm{Na}, i, \text{exact}}\|_{L^2}$ &
    $\|c_{\mathrm{K},  i} - c_{\mathrm{K},  i, \text{exact}}\|_{L^2}$ &
    $\|c_{\mathrm{Cl}, i} - c_{\mathrm{Cl}, i, \text{exact}}\|_{L^2}$ &
    $\|\phi_i - \phi_{i, \text{exact}}\|_{L^2}$ \\
    \noalign{\smallskip}\svhline\noalign{\smallskip}
    1    & \num{9.01e-03}  (--)  & \num{9.00e-03}  (--)  & \num{1.80e-02}  (--)  & \num{9.26e-02}  (--)  \\
    1/2   & \num{2.33e-03} (1.95) & \num{2.33e-03} (1.95) & \num{4.67e-03} (1.95) & \num{2.48e-02} (1.90) \\
    1/4   & \num{5.88e-04} (1.99) & \num{5.88e-04} (1.99) & \num{1.17e-03} (1.99) & \num{6.30e-03} (1.98) \\
    1/8  & \num{1.47e-04} (2.00) & \num{1.47e-04} (2.00) & \num{2.94e-04} (2.00) & \num{1.58e-03} (2.00) \\
    1/16 & \num{3.68e-05} (2.00) & \num{3.63e-05} (2.02) & \num{7.26e-05} (2.02) & \num{3.95e-04} (2.00) \\
    \noalign{\smallskip}\hline\noalign{\smallskip}
    &
    $\|c_{\mathrm{Na}, e} - c_{\mathrm{Na}, e, \text{exact}}\|_{L^2}$ &
    $\|c_{\mathrm{K},  e} - c_{\mathrm{K},  e, \text{exact}}\|_{L^2}$ &
    $\|c_{\mathrm{Cl}, e} - c_{\mathrm{Cl}, e, \text{exact}}\|_{L^2}$ &
    $\|\phi_e - \phi_{e, \text{exact}}\|_{L^2}$ \\
    \noalign{\smallskip}\svhline\noalign{\smallskip}
    1     & \num{3.12e-02}  (--)  & \num{1.04e-02}  (--)  & \num{4.16e-02}  (--)  & \num{6.13e-02}  (--) \\
    1/2     & \num{8.08e-03} (1.95) & \num{2.69e-03} (1.95) & \num{1.08e-03} (1.95) & \num{1.67e-02} (1.88) \\
    1/4    & \num{2.04e-03} (1.99) & \num{6.79e-04} (1.99) & \num{2.72e-03} (1.99) & \num{4.25e-03} (1.97) \\
    1/8   & \num{5.09e-04} (2.00) & \num{1.70e-04} (2.00) & \num{6.78e-04} (2.00) & \num{1.07e-03} (1.99) \\
    1/16  & \num{1.26e-04} (2.00) & \num{4.19e-05} (2.02) & \num{1.68e-04} (2.02) & \num{2.68e-04} (2.00) \\
    \noalign{\smallskip}\hline\noalign{\smallskip}
    \end{tabular}
\end{table}
\begin{table}
    \centering
    \caption{Error norms and convergence rates
    for the three-dimensional verification problem
    on meshes of increasing resolution characterized
    by the edge lengths $h$.
    The coarsest mesh has mesh edge length $h_0$.
    }\label{tab:convergence_rates_3D}
    \begin{tabular}{p{0.9cm}
        >{\centering\arraybackslash}p{2.65cm}
        >{\centering\arraybackslash}p{2.65cm}
        >{\centering\arraybackslash}p{2.65cm}
        >{\centering\arraybackslash}p{2.65cm}
    }
    \hline\noalign{\smallskip}
    $h/h_0$ &
    $\|c_{\mathrm{Na}, i} - c_{\mathrm{Na}, i, \text{exact}}\|_{L^2}$ &
    $\|c_{\mathrm{K},  i} - c_{\mathrm{K},  i, \text{exact}}\|_{L^2}$ &
    $\|c_{\mathrm{Cl}, i} - c_{\mathrm{Cl}, i, \text{exact}}\|_{L^2}$ &
    $\|\phi_i - \phi_{i, \text{exact}}\|_{L^2}$ \\
    \noalign{\smallskip}\svhline\noalign{\smallskip}
    1    & \num{6.70e-03}  (--)  & \num{6.70e-03}  (--)  & \num{1.34e-02}  (--)  & \num{6.82e-02}  (--)  \\
    1/2    & \num{1.79e-03} (1.90) & \num{1.79e-03} (1.90) & \num{3.58e-03} (1.90) & \num{1.93e-02} (1.82) \\
    1/4   & \num{4.54e-04} (1.98) & \num{4.55e-04} (1.98) & \num{9.09e-04} (1.98) & \num{4.99e-03} (1.95) \\
    1/8  & \num{1.14e-04} (2.00) & \num{1.14e-04} (2.00) & \num{2.28e-04} (2.00) & \num{1.26e-03} (1.99) \\
    1/16 & \num{2.83e-05} (2.01) & \num{2.83e-05} (2.01) & \num{5.66e-05} (2.01) & \num{3.16e-04} (2.00) \\
    \noalign{\smallskip}\hline\noalign{\smallskip}
    &
    $\|c_{\mathrm{Na}, e} - c_{\mathrm{Na}, e, \text{exact}}\|_{L^2}$ &
    $\|c_{\mathrm{K},  e} - c_{\mathrm{K},  e, \text{exact}}\|_{L^2}$ &
    $\|c_{\mathrm{Cl}, e} - c_{\mathrm{Cl}, e, \text{exact}}\|_{L^2}$ &
    $\|\phi_e - \phi_{e, \text{exact}}\|_{L^2}$ \\
    \noalign{\smallskip}\svhline\noalign{\smallskip}
    1     & \num{3.55e-02}  (--)  & \num{1.12e-02}  (--)  & \num{4.73e-02}  (--)  & \num{6.74e-02}  (--) \\
    1/2     & \num{9.47e-03} (1.90) & \num{3.16e-03} (1.90) & \num{1.26e-03} (1.90) & \num{1.96e-02} (1.78) \\
    1/4    & \num{2.41e-03} (1.98) & \num{8.02e-04} (1.98) & \num{3.21e-03} (1.98) & \num{5.11e-03} (1.94) \\
    1/8   & \num{6.03e-04} (2.00) & \num{2.01e-04} (2.00) & \num{8.04e-04} (2.00) & \num{1.29e-03} (1.98) \\
    1/16  & \num{1.50e-04} (2.01) & \num{4.99e-05} (2.01) & \num{2.00e-04} (2.01) & \num{3.24e-04} (2.00) \\
    \noalign{\smallskip}\hline\noalign{\smallskip}
    \end{tabular}
\end{table}

\subsection*{Numerical verification with a realistic cerebral tissue geometry}
\begin{figure}
    \centering
    \includegraphics[width=\textwidth]{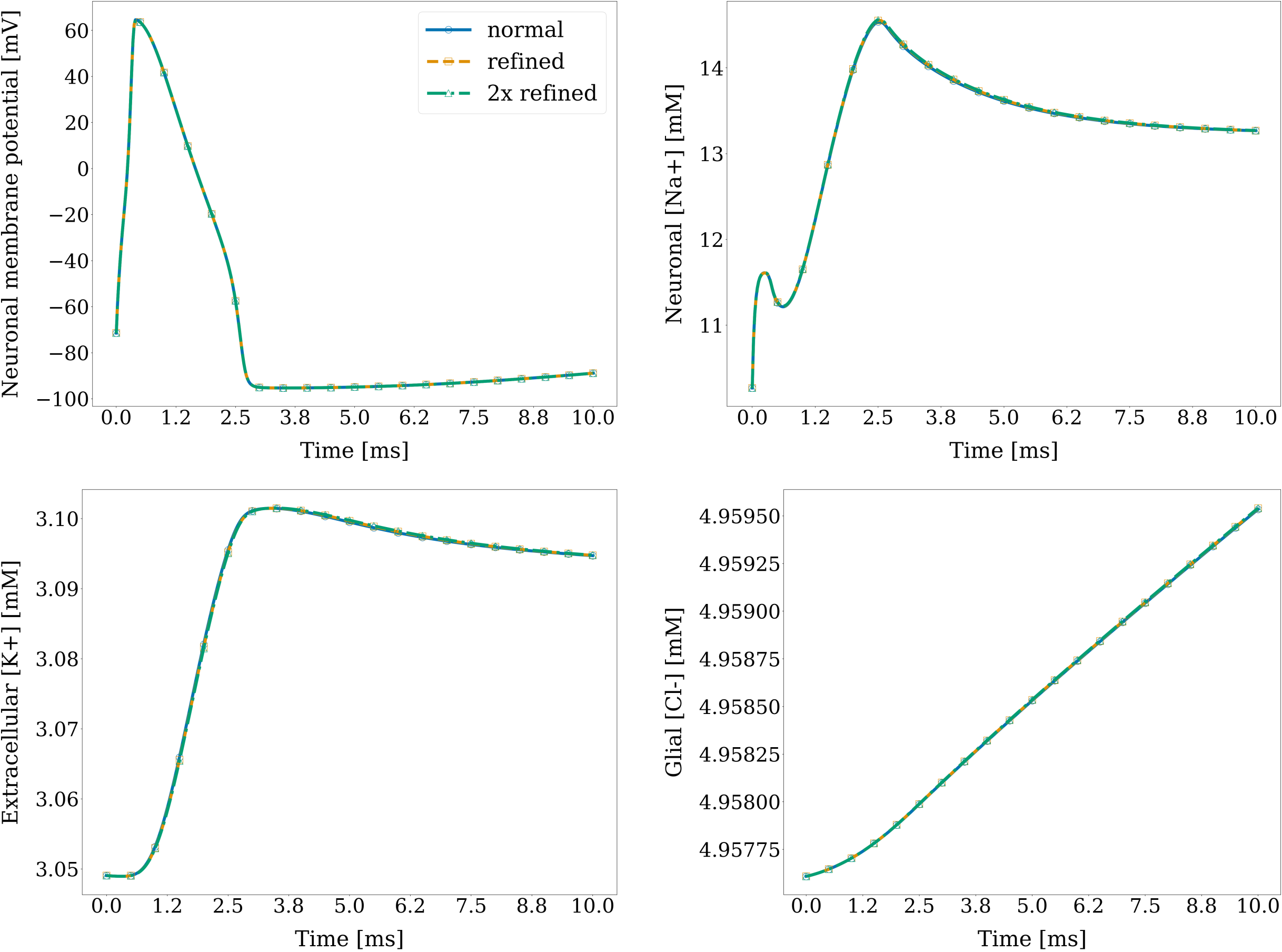}
    \caption{Convergence of solution under mesh refinement on the $L=5$ $\mu$m,
    $N=100$ mouse visual cortex meshes. Three meshes of increasing refinement
    are considered, and we simulate an action potential as described in
    \Cref{subsec:simulating_low_neuronal_activity}.
    The membrane potential is evaluated on the neuronal
    membrane at $\xx = (1186, 3241, 3594)$ nm, while the concentrations are evaluated in the points
    described in~\Cref{fig:physiological_study_setup}.
    Common legend for all subfigures.
    }
    \label{fig:refinement_study}
\end{figure}

With the spatial convergence properties verified on idealized
geometries, we here verify that the numerical solution of the KNP-EMI model converges on the mouse visual cortex tissue mesh. We perform a convergence study 
by solving the KNP-EMI equations on the original $L=5$ $\mu$m, $N=100$ mesh (generated with EMI-Meshing as introduced in~\Cref{sec:reconstruction}),
and two uniformly refined versions of the mesh. 
We run a similar simulation as described in~\Cref{sec:physiological_simulations};
an axon is stimulated with the stimulus current density defined in~\Cref{eq:stimulus_current_density}
with $T_{\mathrm{stim}}=1$ s. The simulation is run for
800 timesteps with $\Delta t = \num{2.5e-5}$ to a final time
of $T = 10$ ms. We evaluate the membrane potential in the point
$\xx = (1186, 3241, 3594)$ nm on the stimulated neuron.
The neuronal intracellular sodium concentration,
the extracellular potassium concentration proximal to the neuron, and
the glial intracellular $\text{Cl}^{-}$ concentration
are all evaluated in the same points
as described in \Cref{sec:physiological_simulations},
see~\Cref{fig:physiological_study_setup}. 
The solutions obtained are similar to the results
observed in~\Cref{sec:physiological_simulations},
and the results agree well across the different mesh versions
(\Cref{fig:refinement_study}).

\end{document}